\documentclass[oldversion]{aa}

\usepackage{graphicx}
\usepackage{txfonts}
\usepackage{amssymb}
\usepackage{longtable}
\usepackage{natbib}
\usepackage{rotating}
\bibpunct{(}{)}{;}{a}{}{,} 

\begin{document}

\titlerunning{A mid-IR study of Hickson Compact Groups II }
\authorrunning{Bitsakis et al.}

\title{A mid-IR study of Hickson Compact Groups\\ II. Multi-wavelength analysis of the complete GALEX-Spitzer Sample}

	\author{T. Bitsakis
		\inst{1}
       \and
     	V. Charmandaris
		\inst{1,2,3}		
       \and
     	E. da Cunha
		\inst{4}		
       \and
     	T. D\'iaz-Santos
		\inst{1}		
	  \and
	E. Le Floc'h
		\inst{5}
	  \and
	G. Magdis
		\inst{6}
}

\institute{Department of Physics \& ICTP, University of Crete, GR-71003, Heraklion, Greece
	  \and
	IESL/Foundation for Research \& Technology-Hellas, GR-71110, Heraklion, Greece
	\and
	Chercheur Associ\'e, Observatoire de Paris, F-75014,  Paris, France
	\and
	Max Planck Institute f\"{u}r Astronomie, D-69117, Heidelberg, Germany
	\and
	Laboratoire AIM, CEA/DSM - CNRS - Universit\'e Paris Diderot, IRFU/Service d'Astrophysique,  F-91191, Gif-sur-Yvette Cedex, France
	\and
	University of Oxford, Department of Physics, Keble Road, Oxford OX1 3RH, UK
}

\offprints{T. Bitsakis,  e-mail: bitsakis@physics.uoc.gr}

\abstract{We present a comprehensive study on the impact of the environment of compact galaxy groups on the evolution of their members 
using a multi-wavelength analysis, from the UV to the infrared, for a sample of  32 Hickson compact groups (HCGs) containing 135 galaxies. 
Fitting the SEDs of all galaxies with the state-of-the-art model of da Cunha (2008) we can accurately calculate their mass, SFR, and 
extinction, as well as estimate their infrared luminosity and dust content. We compare our findings with  samples of field galaxies, 
early-stage interacting pairs, and cluster galaxies with similar data.  We find that classifying the groups as dynamically ``old'' 
or ``young'', depending on whether or not at least one quarter of their members are early-type systems, is physical and consistent with 
past classifications of HCGs based on their atomic gas content. Dynamically ``old'' groups are more compact and display higher velocity 
dispersions than  ``young'' groups. Late-type galaxies in dynamically ``young'' groups have specific star formation rates (sSFRs), NUV-r, 
and mid-infrared colors which are similar to those of field and early stage interacting pair spirals. Late-type galaxies in dynamically 
``old'' groups have redder NUV-r colors, as they have likely experienced several tidal encounters in the past building up their stellar 
mass, and  display lower sSFRs. We identify several late-type galaxies which have sSFRs and colors similar to those of elliptical galaxies, 
since they lost part of their gas due to numerous interactions with other group members. Also, 25\% of the elliptical galaxies in these 
groups have bluer UV/optical colors than normal ellipticals in the field, probably due to star formation as they accreted gas from other 
galaxies of the group, or via merging of dwarf companions. Finally, our SED modeling suggests that in 13 groups, 10 of which are dynamically ``old", 
there is diffuse cold dust in the intragroup medium. All this evidence point to an evolutionary scenario in which the effects of the group 
environment and the properties of the galaxy members are not instantaneous. Early on, the influence of close companions to group galaxies 
is similar to the one of galaxy pairs in the field. However, as the time progresses, the effects of tidal torques and minor merging, shape the 
morphology and star formation history of the group galaxies, leading to an increase of the fraction of early-type members and a rapid built 
up of the stellar mass in the remaining late-type galaxies.}


\maketitle

\section{Introduction}

It has become increasingly evident that interactions and  merging of galaxies have contributed substantially to their evolution, both
 in terms of their stellar population as well as their morphological appearance. Compact galaxy groups, with their high galaxy density and 
signs of tidal interactions among their members, are ideal systems to study the impact on environment to the evolution of galaxies. 
The Hickson Compact Groups (HCGs) are 100 systems of typically 4 or more galaxies in a compact configuration on the sky  \citep{Hickson82}. 
They contain a total of 451 galaxies and  are mostly found in relatively isolated regions where no excess of surrounding galaxies can be 
seen, reflecting a strong local density enhancement. The HCGs occupy a unique position in the framework of galaxy evolution, bridging the 
range of galaxy environments, from field and loose groups to cores of rich galaxy clusters. 
The fact that the original selection of the HCG members did not include redshift information, led to the inclusion of interlopers among 
them, the most famous being NGC 7320 in  Stephan's Quintet (HCG  92).  This led to a debate as to whether compact groups  are line-of-sight 
alignments of galaxy pairs within loose groups, or filaments seen end-on \citep{Mamon86, Hernquist95}. However, the detection of hot X-ray 
gas in $\sim$75\% of the HCGs by \citet{Ponman96} implies that they reside in a massive dark matter halo and thus are indeed physically 
dense structures. Numerical simulations indicate that in the absence of velocity information, raising the minimum surface brightness criterion 
for the group used by Hickson would help eliminate  interlopers \citep[see][]{McConnachie08}.

Because of the nature of these groups, the high density enhancements in addition to the low velocity dispersions ($\sim$250 km s$^{-1}$), 
make them ideal to study the effects of galaxy interactions. \citet{Hickson82} found that the majority of HCGs display an excess of elliptical 
galaxies, $\sim$31\% of all members compared to the field, while the fraction of spiral galaxies and irregular is only 43\%, nearly a factor 
of two less of what is observed in the field. Optical imaging by \citet{Mendes94} showed that  43\% of all HCG galaxies display morphological 
features of interactions and mergers, such as bridges, tails and other distortions. Similar indications of interactions are seen in maps of 
the atomic hydrogen distribution in selected groups by \citet{Verdes01}. Moreover, \citet{Hickson89} found that the fractional distribution of 
the ratio of far-infrared (far-IR) to optical luminosity in HCG spiral galaxies is significantly larger than that of isolated galaxies, 
suggesting that for a given optical luminosity, spiral galaxies in groups have higher infrared luminosities. Comparison of HCG spirals with 
those in clusters of galaxies from \citet{Bicay87} reveals that the distributions of the IR to optical luminosity, as well as the 60 to 
100$\mu$m far-IR color are similar. Finally, nuclear optical spectroscopy studies indicate that almost 40\% of the galaxies within these groups 
display evidence of an active galactic nucleus (AGN, \citealt{Martinez10,Shimada00}). All these clues are consistent with an evolutionary 
pattern where tidal encounters and the accretion of small companions by the group members, redistribute the gas content of the groups and 
affect the morphology of their members.

\citet{Verdes01} and \citet{Borthakur10} have proposed an evolutionary sequence for the HCGs based on the amount and spatial distribution of 
their neutral atomic gas. Using HI maps they classified the groups  into three phases based on the ratio of the gas content within the 
galaxies over the total observed in the group. According to their scenario, a loose galaxy group starts to contract under the gravity to 
form a more compact one. During this first phase the HI gas is still mostly found in the individual galaxies. Then as the group evolves, it 
enters the second phase and a fraction of the atomic hydrogen is extracted from the galaxy disks into the intragroup medium, probably due to 
tidal forces, and part of it becomes fully ionized. Finally, in the third phase, the dynamical friction leads to a decrease in the separation 
between the group members and the group becomes more compact. More than  80\% of the HI originally in the disks of the galaxies has been 
displaced. Some of it is seen redistributed in a common envelope surrounding all groups members, and a fraction has likely been converted 
into molecular gas fueling star formation and/or accretion onto an AGN. A similar classification was proposed by \citet{Johnson07} and  
\citet{Tzanavaris10}. These authors separated the groups using the ratio of the M$_{\rm HI}$ over the dynamical mass of the group, the 
so-called ``HI richness", and found that it is correlated to the specific star formation (sSFR) of their galaxies. They also found that 
galaxies in gas-poor groups display colors representative of a normal stellar population and observed a bimodality in the mid-infrared colors, 
as well as in the sSFRs of their members. They suggested that this is caused by enhanced the star formation activity which lead the galaxies 
in groups to evolve faster. \citet{Walker09} had concluded that a similar bimodality in the mid-IR is also observed in the colors of galaxies 
in the Coma Infall region.

A necessary step to determine the evolutionary state of HCGs, is the analysis of not just the morphology of the group members, but of 
their stellar population and star formation history. In our first paper (\citealt{Bitsakis10}), we commenced exploring the properties of an 
initial sample of 14 HCGs in  mid-IR wavelengths and found that a large fraction of the early type galaxies in groups, have mid-IR colors 
and spectral energy distributions (SEDs) consistent with those of late-type systems. We suggested that this possibly stems from an 
enhanced star formation, as a result of accretion of  gas rich dwarf companions. We found no evidence that the star formation rate (SFR) and 
built up of stellar mass of late-type galaxies in groups is different from what is observed in early-stage interacting pairs, or spiral 
galaxies in the field. However, when we separated the groups according to the fraction of their early type 
members, they appear to differentiate from the control samples. This would  suggests an evolutionary separation of HCGs and  
can provide a better insight on the nature of these groups. 

Despite the progress in the analysis of the properties of these groups, there are still several open questions. Is the bimodality 
of the mid-IR colors and sSFRs indeed linked with the evolutionary sequence of these galaxies, or it is also observed in other galaxies? How 
do the properties of HCG galaxies compare to those of galaxies in other environments? Is it really physical to classify the evolutionary 
stages of these groups also according to the fraction of their early-type systems? Is this classification meaningful in terms of galaxy 
properties and colors and does it agree with the other classifications? To answer  these questions we need  multi-wavelength data, as well 
as a theoretical model to fit their global SED, in order to obtain the best possible constraints on the physical parameters. In this paper 
we present the first such analysis for a large sample,  nearly one third of all Hickson groups, for  which we have retrieved observations 
from the  UV to the infrared part of the spectrum.  

The structure of the paper is the following. We present our samples, the observations and the data reduction in Section 2. In Section 3, 
we describe the model used to fit the data as well as the basic physical parameters we can derive, along with their uncertainties. Our 
results are shown in Section 4, and our conclusions are summarized in Section 5. In an Appendix we provide additional information on10 
early-type galaxies which have peculiar mid-IR colors and SEDs.

\section{Observations and Data Reduction}

Our sample was constructed from the original Hickson (1982) catalogue of 100 groups, using  as criterion the availability of 
high spatial resolution 3.6 to 24$\mu$m mid-infrared imagery from the Spitzer Space Telescope archive, as well as UV imaging 
from GALEX. The infrared data are essential to probe the properties of the energy production in nuclei 
of galaxies, some of which may be enshrouded by dust, while the  UV is necessary to properly estimate the effects 
of extinction and accurately account for the global energy balance when we model their SED. These constrains resulted in a 
sample of 32 compact groups  containing 135 galaxies, 62 (46\%) of which are early-type (E's \& S0's) and 73 (54\%) are 
late-type (S/SB's \& Irr's). This nearly triples the sample of 14 HCGs we studied in \citet{Bitsakis10}. We  verified that 
all galaxies of the groups are sufficiently separated in order to be able to obtain accurate photometry from the UV to the 
mid-IR, and that the average group properties such as type of galaxy, stellar mass, star formation rates are representative 
of the whole Hickson sample. We should note that 7 groups contain interlopers along the line of sight. For these groups the 
number of physical members is three, below the lower limit of four members introduced by \citet{Hickson82}.  In Table~\ref{tab:sample} we present a summary of all observations available. The complete photometry is presented in Table~\ref{tab:flux}.

\subsection{GALEX data}

The UV data presented in this paper were obtained from the Galaxy Evolution Explorer (GALEX) All Sky Survey (AIS),  the 
Medium Imaging Survey (MIS), as well as from Guest Investigator Data data publicly available in the GALEX archive. GALEX is a 
50cm diameter UV telescope that images the sky simultaneously in both FUV and NUV channels,  centered at $1540\AA{}$ and 
$2300\AA{}$, respectively. The field of view (FOV) is approximately circular  with a diameter of $1.2^{o}$ and a resolution 
of about 5.5'' (FWHM) in the NUV. The data sets used in this paper are taken from the GALEX sixth data release (GR6). More 
technical details about the GALEX instruments can be found  in \citet{Morrissey05}. We performed aperture photometry and 
carefully calculated the isophotal contours around each source to account for variations in the shape of the 
emitting region, since most of the sources have disturbed morphologies. By examining the local overall background  for each 
galaxy, we defined a limiting isophote 3$\sigma$ above it and we measured the flux within the region, after  subtracting 
the corresponding sky. Finally, the conversion from counts to UV fluxes was done using the conversion coefficients given 
in the headers of each file. Our measurements are presented in Table~\ref{tab:flux}. 

\subsection{SDSS-Optical data}

We compiled the B and R band images of all the galaxies in our sample as reported in Table 2 of \citet{Hickson89}, who had 
observed the groups for 200s in both B and R bands using FOCAS1 on the Canada French Hawaii telescope resulting in images 
of  0.42'' pixel$^{-1}$ and a typical resolution of 1.2'' (FWHM).  In addition, imaging data of 74 galaxies in the $u, g, 
r, i, z$ bands, centered at 3557, 4825, 6261, 7672 and 9097$\AA{}$ respectively,  were recovered from the Sloan Digital 
Sky Survey (SDSS). We used the SDSS DR7 ``model magnitudes'', which reflect the integrated light from the whole galaxy and 
are best suited for comparisons with total photometry from ultraviolet to infrared wavelengths.

Furthermore, optical spectra were available for 94 galaxies, 70\% of our sample. Using the BPT diagram, \cite{Martinez10} 
classified the nuclei of 67 galaxies in our sample, while the remaining were obtained from \citet{Shimada00}, \citet{Hao05} 
and \citet{Veron06}. The results are shown in Table~\ref{tab:flux}. Galaxies without emission are referred as ``unclas.'' 
(unclassified), as they display only stellar absorption features, ``HII'' are the galaxies with a starburst nucleus, ``AGN'' 
(or LINER \& Sy2) are the ones with an active nucleus. A number of galaxies are classified as transition objects (TO) since 
their emission line ratios are intermediate between \citealt{Kauffman03} and \citealt{Kewley06} criteria. Based on these 
results for the galaxies in our sample where nuclear classification was available, 37\% host an AGN, which is close 
to the 40\% which \citet{Martinez10} and \citet{Shimada00} found for their samples.

\subsection{Near-Infrared Observations}
Deep near-IR observations were obtained for 15 of the groups, using the Wide Field Infrared Camera (WIRC) of the 5m  Hale 
telescope at Palomar. As we discussed in  \citet{Bitsakis10} all groups were imaged in the J, H, and Ks bands for an on-source 
time of 20 minutes per filter \citep{Slater04}. Source extraction was performed with SExtractor \citep{Bertin96}. Our 
1$\sigma$ sensitivity limit was $\sim$ 21.5 mag arcsec$^{-2}$ in J and H bands and $\sim$20.5 mag arcsec$^{-2}$ for Ks. For 
the remaining 17 groups of our sample, the near-IR fluxes were obtained from the Two Micron All Sky Survey (2MASS; 
\citealt{Skrutskie06}). In cases where the proximity of galaxy pairs was affecting the reliability of the 2MASS photometry, 
we used the reduced 2MASS images and performed aperture photometry ourselves defining the same aperture we used in other 
wavelengths. However, for HCG6  the 2MASS photometry was problematic and we observed the group using the wide-field near-IR 
camera of the1.3m telescope at Skinakas Observatory, in Crete (Greece). The observations were carried out between 21 and 23 
of September 2010. The group was imaged in the J, H and Ks bands for an on-source time of 30min per filter and flux calibration 
was performed using the a set of near-IR standard stars. Data reduction and aperture photometry was performed using IDL 
specialized routines. All near-IR photometry for our sample is available in Table~\ref{tab:flux}.

\subsection{Mid-Infrared Spitzer Observations}

Mid-IR observations for 11 of the groups were performed by us, using Spitzer Space Telescope between 2008 January and 2009 
March. We used the Infrared Array Camera \citep[IRAC,][]{Fazio04} and obtained 270sec on-source exposures with the $3.6, 
4.5, 5.8$ and $8.0 \mu$m broadband filters. Each group was also imaged with the Multiband Imaging Photometer for Spitzer 
\citep[MIPS,][]{Rieke04} with a 375.4sec on source exposure at $24 \mu$m.  Details on the analysis of these data were 
presented in \citet{Bitsakis10}.  The Spitzer mid-IR fluxes for 8 more groups of the sample were obtained from \citet{Johnson07}. 
In brief, the IRAC observations were taken in high dynamic range with 270 or 540sec exposures 
depending on the group. MIPS 24$\mu$m images were obtained by the same authors with exposures of 180 or 260sec duration for 
the same reason. Finally, for 13 groups we recovered the IRAC and MIPS $24 \mu$m data from the Spitzer archive (PIDs: 159; 
198; 50764; 40385) and reduced them as in  \citet{Bitsakis10}. The details of these observations are presented in 
Table~\ref{tab:obs}, and the compilation of all mid-IR measurements is included in Table~\ref{tab:flux}.

\subsection{Far-Infrared data}

Integrated far-IR photometry at 60 and 100$\mu$m data for each group as a whole, was obtained from \citet{Allam96}. However, 
because  of the compact environment and small angular separation of the galaxies in the HCGs, the coarse angular resolution of 
IRAS at the 60 and 100$\mu$m ($\sim$100''), enabled us to resolve only 31 individual galaxies of our sample which were sufficiently 
isolated and bright.

Using the recently released AKARI data  we retrieved far-IR observations for 26 galaxies of our sample (22 of them were common with 
the IRAS sample), which were obtained with the Far-IR Surveyor (FIS) and processed with the AKARI official pipeline software version 
20080530 \citep{Okada08}. The photometry using N60, WIDE-S, WIDE-L, and N160 filters, centered at 65, 90, and 140 160 $\mu$m was 
downloaded from the AKARI on-line archive and is reported along  with the IRAS data on Table~\ref{tab:flux}. We should note that 
the FWHMs of the AKARI PSF is $\sim$45'' at  65 and 90$\mu$m and $\sim$60'' at 140 and 160$\mu$m. Even though this is better than 
IRAS, as we will discuss in more detail later, it places restrictions in interpreting the far-IR properties of the groups and 
the spatial distribution of their cold dust content. 

\subsection{Comparison Samples}

In order to put the properties and evolution of the group galaxies into context, we must compare their observables and derived 
physical parameters with other control samples. Since dynamical interactions are the main drivers of galaxy evolution, we examined 
isolated field galaxies,  as well as  systems which are dynamically interacting, such galaxy pairs and galaxies found in clusters 
for which we could obtain data of similar wavelength coverage. The samples we used in our analysis are the following:

\subsubsection{Isolated Field Galaxies}

A well studied sample with superb data coverage is the 75 ``normal",  mostly isolated field galaxies from the Spitzer Infrared 
Nearby Galaxies Survey  \citep[SINGS;][]{Kennicutt03, Dale05}. We should note that the SINGS sample was selected explicitly to 
cover a wide range in Hubble type and luminosity and as a result it is not characteristic of a volume or flux limited population. 
Most objects are late-type systems that have angular sizes between 5$'$ and 15$'$. It also contains four early-type galaxies 
(NGC\,855, NGC\,1377, NGC\,3773, and NGC\,4125) with the last one being a LINER. Since the SINGS sample does not have many early type galaxies, 
we used the 9 isolated early-type galaxies with available mid-IR imaging described in \citet{Temi04} and expanded it with another 
4 isolated galaxies (NGC1404, NGC1199, NGC5363 \& NGC5866) for which Spitzer mid-IR imaging was also available.

\subsubsection{Interacting Galaxy Pairs}

This sample is drawn from the 35 nearby (v$<$11.000 km\,s$^{-1}$) early stage interacting galaxy pairs of \citet{Smith07a}. The 
galaxies are tidally disturbed and fairly extended, having  linear sizes $>3'$. For the purposes of this work only 26 pairs from 
the initial sample were used, in which all the data were available. As discussed in \citet{Bitsakis10}, the stellar mass 
distributions are similar between this sample as well ours and the SINGS.

\subsubsection{Field, Group, and Cluster galaxies}

To compare in more detail the UV-optical colors of the galaxies in HCGs with the colors of galaxies found in the field, as well 
as in other groups and clusters, we used the volume limited sample of \citet{Haines08}. This sample contains 1994 galaxies in redshift 
range 0.005$\leq$z$\leq$0.037, selected by cross-correlating the SDSS-DR4 sample with the GALEX GR3 photometric catalogues. Using 
the  H$\alpha$ equivalent widths of M$_{r}$$<$-18 galaxies the authors were able to separate the galaxies into passively evolving 
and star-forming, having EW(H$\alpha$)$<$2 and EW(H$\alpha$)$>$2 respectively. To quantify the effects of local mass density and 
environment the authors calculated the local galaxy number density, $\rho$, in units of  projected area, in Mpc$^{2}$, around a 
central galaxy and within a radial velocity bin of 500km\,s$^{-1}$. They found that for $\rho < $\,0.5Mpc$^{-2}$(500\,km\,s$^{-1}$)
$^{-1}$ they can produce a pure field sample. If a galaxy has $\rho > $\,4.0Mpc$^{-2}$(500\,km\,s$^{-1}$)$^{-1}$ there is a 
90\% probability that it lies within the virial radius of a group or a cluster.

\section{Estimating the physical parameters of the galaxies}

\subsection{Fitting the UV, Optical, and IR SEDs}

We use the state-of-the-art physically motivated model of \citet{daCunha08}\footnote{The \citet{daCunha08}  model is publicly available 
at http://www.iap.fr/magphys. }, to fit the SEDs of the galaxies in our sample.  As 
discussed in detail by \citet{daCunha08} the model assumes that the source of energy in a galaxy is due to an ensemble of stellar 
populations (no AGN heating is included) whose emission is partially absorbed by dust and re-emitted at longer wavelengths. The 
model treats the complete SED from the UV to the far-IR and allows us to derive not simply best fit physical parameters to the 
data, but also to provide the range of their median-likelihood values  which are consistent with the observations. To achieve this, 
the model adopts a Bayesian approach which draws from a large library of random models encompassing all plausible parameter 
combinations, such as star formation histories, metallicities, dust optical depths as well as dust masses and temperatures. Clearly 
the wider the wavelength coverage, the more robust the derived parameters. Thus to properly estimate the effect of dust extinction, the UV 
and optical range needs to be well sampled, while to estimate dust masses and luminosities mid- and far-IR coverage is essential. 
In addition to the HCG sample, we used the model to fit the interacting pairs of \citet{Smith07a} and  the results are presented in 
the Table~\ref{tab:intpairs}. The galaxies of the SINGS sample had already been fit by \citet{daCunha08}. 

\begin{figure}
\begin{center}
\resizebox{\hsize}{!}{\includegraphics{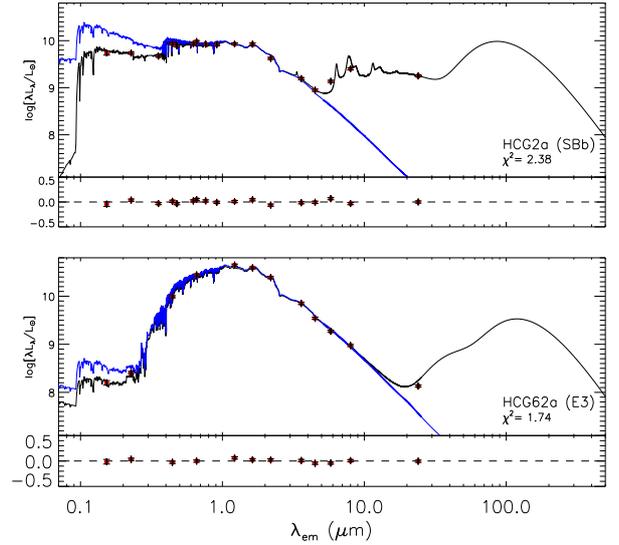}}
\caption{Example of best fit models (in black line) to the observed spectral energy distributions  
of two galaxies of our sample. One is a typical late-type galaxy (HCG2a; top panel) and the second 
is a quiescent elliptical galaxy (HCG62a; bottom panel). In each panel, the blue line shows the 
unattenuated stellar spectrum estimated by the model. The red circles are the observed broadband 
luminosities with their errors as vertical bars. The residuals, 
$(L_\lambda^\mathrm{obs}-L_\lambda^\mathrm{mod})/ L_\lambda^\mathrm{obs}$, are shown at the bottom 
of each panel.  Note the uncertainty in the far-IR shape of the SED due to lack of high spatial resolution data.}
\label{fig:seds}
\end{center}
\end{figure}

\subsubsection{Description of the model}

The \citet{daCunha08} model computes 
the emission by stars in galaxies using the latest version of the \citet{Bruzual03} population synthesis code. This code predicts 
the spectral evolution of stellar populations in galaxies from far-ultraviolet to infrared wavelengths and at ages between 
1$\times$10$^{5}$ and 1.37$\times$10$^{10}$ yr, for different metallicities, initial mass functions (IMFs) and star formation 
histories. In this work, we adopt the \citet{Chabrier03} Galactic-disc IMF.  The model does not take in account the 
energy contribution of an active nucleus to the global SED. Since nearly 40\% of our sources host an optically identified AGN one
could consider that this could introduce a bias in our analysis. However, based on a number of mid-IR diagnostics the influence of the active nucleus to the total infrared emission of the majority of our galaxies is insignificant, thus this limitation does affect seriously our conclusions (see details in Section 4.5). 

The emission from stars is attenuated using a two component dust model of \citet{Charlot00}. This model uses an ``effective 
absorption'' curve for each component. They use this prescription to compute the total energy absorbed by dust in the birth 
clouds (BC) and in the ambient interstellar medium (ISM); this energy is re-radiated by dust at infrared wavelengths. They 
define the total dust luminosity re-radiated by dust in the birth clouds and in the ambient ISM as L$^{BC}_{d}$ and L$^{ISM}_{d}$, 
respectively. The total luminosity emitted by dust in the galaxy is then L$^{tot}_{d}=$L$^{BC}_{d}+$L$^{ISM}_{d}$. The value 
of L$^{BC}_{d}$ and L$^{ISM}_{d}$ is calculated over the wavelength range from 3 to 1000 $\mu$m using four main components:
\begin{itemize}
 \item The emission from polycyclic aromatic hydrocarbons (PAHs; i.e. mid-infrared emission features).
 \item The mid-infrared continuum emission from hot dust with temperatures in the range 130-250 K.
 \item The emission from warm dust in thermal equilibrium with adjustable temperature in the range 30-60 K.
 \item The emission from cold dust in thermal equilibrium with adjustable temperature in the range 15-25 K.
\end{itemize}

A detail analysis of \citet{daCunha08} suggests that the above minimum number of components is required to account for the 
infrared spectral energy  distributions of galaxies in a wide range of star formation histories. 

\subsubsection{Model library}

We use a large random library of star formation histories and dust emission models presented in da \citet{daCunha08}. In this 
library, each star formation history is parameterized using a star formation rate that is exponentially declining with time, 
on top of which random bursts are superimposed. The metallicities of these models are sampled uniformly between 0.2 and 2 times 
solar, and the model ages are distributed uniformly between 0.1 and 13.5 Gyr. Each stellar emission model computed using these 
star formation histories, ages and metallicities is then attenuated by dust using the two-component description of \citet{Charlot00} 
described above, using a wide range of V-band dust optical depths in the birth cloud and ISM component. Each of these model spectra 
are then consistently connected to dust emission spectra spanning a wide range in dust temperatures and fractional contributions of 
each dust emission component to the total dust luminosity. In particular, dust emission models with the same dust luminosity and the 
same relative contributions to total dust luminosity from the birth cloud and ISM (i.e. L$_{d,tot}$ and f$\mu$) are assigned to each stellar 
emission model, as detailed in \citet{daCunha08}.

For each ultraviolet to infrared model spectrum in this library, we compute the synthetic photometry in the GALEX FUV and NUV, SDSS 
ugriz, B, R, near-IR, Spitzer IRAC and MIPS-24, AKARI/FIS and IRAS 60 and 100$\mu$m bands at the redshifts of our galaxies to allow for 
a direct comparison between the observed and model broad-band SEDs as detailed below.

\subsubsection{Spectral fits}

We compare the observed spectral energy distributions of the galaxies in our HCGs to every model in the stochastic library of models 
described above by directly comparing the observed and model broad-band fluxes and computing the $\chi^{2}$ goodness of fit for each model 
in the library. This allows us to compute, for each galaxy in the sample, the full likelihood distributions of several model parameters, 
such as the star formation rate, stellar mass, V-band optical depth in the birth clouds and ISM, dust luminosity etc. We take our final 
estimate of each parameter to be the median of the likelihood distribution, and the associated confidence interval to be the 16th -- 84th 
percentile range of that distribution.

As examples of our fits, we show in Fig.~\ref{fig:seds} the best-fit spectral energy distributions of two galaxies from our sample: one 
quiescent, moderately dusty galaxy (HCG48a, bottom panel); and one actively star-forming, relatively dusty galaxy (HCG7a, top panel).

We note that, as discussed in Sect. 2, even though far-IR measurements are available for each HCG as a whole, most of the galaxies 
are not independently resolved in the far-IR, due to the large IRAS beam sizes. This presents a serious limitation to the sampling of 
our SEDs and consequently to the constraints on the dust temperature, total dust mass and luminosity of each galaxy. One of the strengths 
of this approach is that, by building the full likelihood distributions of the model parameters, we are able to take these uncertainties 
caused by the lack of far-IR observations into account (see also \citealt{daCunha08}).  

\begin{figure}
\begin{center}
\resizebox{\hsize}{!}{\includegraphics{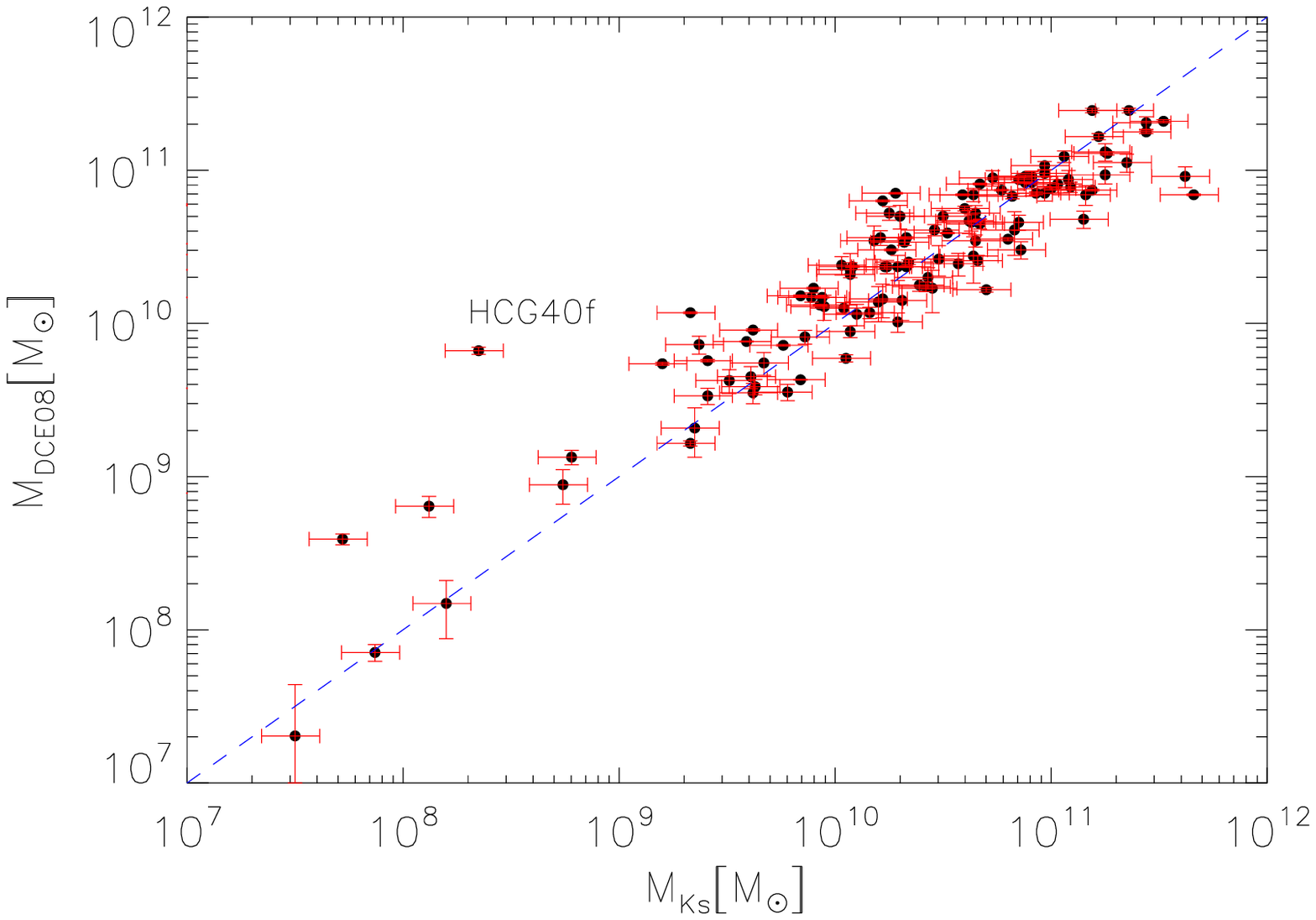}}
\caption{Stellar masses based on the \citet{daCunha08} model (hereafter M$_{\rm DCE08}$) are 
plotted as a function of the  ones derived from \citet{Bell03} relation based on Ks luminosity 
(M$_{\rm Ks}$) after applying the corrections discussed in the text. Vertical error bars 
indicate the 16-84 percentile ranges in the recovered probability distributions while the horizontal 
error bars are the 30\% errors of \citet{Bell03} due to uncertainties in the star formation 
history and dust. The blue dashed line represents the one-to-one relation.}
\label{fig:masses}
\end{center}
\end{figure}

\subsection{Comparison of empirical to model results}

Given the available observations, the stellar mass and star formation rate (SFR) of a galaxy are the two physical parameters 
which are constrained more accurately using our SED modelling. Since in \citet{Bitsakis10} we did not had the multi-wavelength data 
and the model, we used semi-empirical methods, relying on the K-band luminosity and the 24$\mu$m emission, to estimate the stellar mass 
and the SFR respectively. In this section, we compare the model derived parameters with those from the empirical methods, to better 
understand the uncertainties in the interpretation of the results. We also compare them with the methods of \citet{Salim07} and 
\citet{Iglesias06}, which use the NUV and FUV bands of GALEX, respectively, to estimate the SFR, in order to evaluate their consistency 
with the model results. 

The stellar mass of a galaxies can be estimated using the \citet{Bell03} prescription which was calibrated using a large 
sample of galaxies in the local universe. Their formula, based on the K-band luminosity, is:
\begin{equation}
\rm{
M(M_{\odot})=10^{a + b (B-R)}\times L_{Ks}(L_{Ks,\odot})
} 
\end{equation}
where B and R are the B- and R-band magnitudes of the galaxy, L$_{\rm Ks}$ its Ks-band luminosity in units of solar 
luminosities (L$_{\rm Ks,\odot}=4.97\times10^{25}$W), {\it a}=-0.264, and {\it b}= 0.138, with systematic errors of $\sim$30\% due 
to uncertainties in the star formation history and dust. Because these authors used ``diet'' Salpeter IMF, we corrected 
it to a Chabrier IMF (which our model uses). Finally, we applied the color correction described in \citet{Zibetti09} for the 
Charlot \& Bruzual 2007 libraries (CB07), where a is -1.513 and b is 0.750 for the same colors. As we can see in 
Fig.~\ref{fig:masses}, the masses estimated by the model (M$_{model}$) agree with the \citet{Zibetti09} ones. This was somewhat 
expected as the latter recipe was calibrated using the same models. However, there are some outliers such as HCG40f, for which the model 
was not able to  estimate very well the stellar mass since there were not enough observations to well constrain its SED. 

\begin{figure}
\begin{center}
\resizebox{\hsize}{!}{\includegraphics{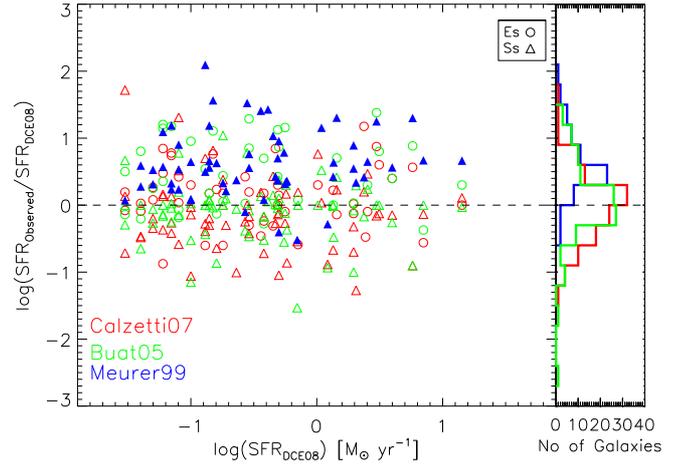}}
\caption{A plot of the ratio of the SFRs derived from the observed FUV or 24$\mu$m fluxes 
over the SFRs derived from the \citet{daCunha08} model as a function of the  model SFR. Red 
open circles and red open triangles are the SFRs derived from the MIPS 24$\mu$m fluxes for 
early and the late-type galaxies respectively. Green open circles and green open triangles 
mark the estimates based on the FUV to infrared excess as described by \citet{Buat05}. 
Filled blue triangles indicate late-type galaxies for which the SFRs are derived from the 
FUV after estimating the extinction using the $\beta$ slope method of \citet{Meurer99}. The 
dashed line is the one-to-one relation. The histogram of the ratios for each method is indicated 
on the right panel.}
\label{fig:sfrs}
\end{center}
\end{figure}

The rate by which a galaxy forms stars, is one of its most important properties, especially in interacting systems which 
evolve very rapidly as they consume the available gas. In the absence of narrow band imaging of the hydrogen recombination 
lines (i.e. H$\alpha$), two of the most common ways to estimate the SFR rely on prescriptions using their UV broad band emission 
from GALEX or the 24$\mu$m thermal dust emission from Spitzer/MIPS. 

Since a significant fraction of the bolometric luminosity of a galaxy is absorbed by interstellar dust 
and re-emitted  in the thermal IR, mid-IR observations can probe the dusty interstellar medium and dust-enshrouded 
SFRs. The efficiency  of the IR luminosity as a SFR tracer depends on the contribution of young stars to heating 
of the dust and the optical depth of the dust in the star forming regions.  \citet{Calzetti07} using a wealth of 
multi-wavelength observations for 33 SINGs galaxies provided the following recipe:
\begin{equation}
\rm{
SFR_{24}(M_{\odot}yr^{-1})=1.27\times10^{-38}(L_{24}(erg\,s^{-1}))^{0.885}
}
\end{equation}
to calculate the SFR from the 24$\mu$m luminosities obtained with Spitzer. We used this formula to estimate the SFR of 
the galaxies in our sample and display them as red points in Fig.~\ref{fig:sfrs}. As we can see the 24$\mu$m estimates 
and those from the model fit to the whole SED, agree fairly well, having a ratio with a median value of unity and a 
standard deviation of 0.51dex. The scatter does not seem to correlate with the galaxy mass or luminosity. To examine whether 
the scatter is due to the absence of far-IR measurements, for a large fraction of the galaxies in our sample we selected 
35 galaxies for which IRAS and Akari fluxes were available and fit them again with the model, removing this time the 
far-IR points. We then compared the new estimates of the SFRs with the ones when all data were used and 
found that they were very similar, their average ratio being 0.95 and the corresponding standard deviation 0.19. 
A possible explanation for the scatter between the results of the model and the \citet{Calzetti07}  method could be 
related to variations in the metallicity of the galaxies \citep[i.e.][]{Rosenberg08}. Another factor which may influence 
the SFR values, related to the star formation history of the galaxies, is the fact that in the \citet{daCunha08} model 
the star formation rates are averaged over 100Myr, while the \citet{Calzetti07} assumes a timescale of 10Myr. 

In the UV the integrated spectrum is dominated by young stars, and so the SFR scales linearly with luminosity. However in 
order to use the UV to trace SFR, we have to correct the attenuated UV fluxes for dust extinction. One method to do this 
is to use the $\beta$-slope correction, 

\begin{equation}
 \beta(GALEX)={log(f_{FUV})-log(f_{NUV}) \over log(\lambda_{FUV})-log(\lambda_{NUV}))}
\end{equation}
where f$_{FUV}$ and f$_{NUV}$ are the flux densities per unit wavelength in the FUV and NUV bands respectively,  
$\lambda_{FUV}=1520$\AA, and $\lambda_{FUV}=2310$\AA. Then we can apply the relation of \citet{Meurer99} for starburst 
galaxies, 
\begin{equation}
\rm L_{\rm FUV,cor} = 10^{0.4(4.43+1.99\beta)}L_{\rm FUV,obs} 
\end{equation}
where  L$_{\rm obs}$ and L$_{\rm cor}$ are the observed and extinction corrected  FUV luminosities and finally we can derive the SFRs 
using the relation of \citet{Salim07}:  
\begin{equation}
\rm{
SFR_{FUV}(M_{\odot}yr^{-1})=6.84\times10^{-29} \, L_{FUV,cor} 
}
\end{equation}
for Chabrier IMF. 

The SFRs of the early-type galaxies cannot be estimated using this method since the $\beta$-slope 
 is mainly determined by the old  stellar population rather than the dust extinction and thus their extinction 
corrected UV luminosities are completely overestimated. Even for  late-type galaxies though, when we compare in Fig. 3 
the FUV estimates of the SFR with the corresponding values obtained with the model we find that the former are overestimated 
by a factor of 0.56dex with a global scatter of 0.46dex. We attribute this to the calibration of the extinction correction of 
\citet{Meurer99} who used a sample of local UV-selected starburst galaxies with high dust content, quite dissimilar to the HCG 
galaxies. As discussed in detail by \citet{Kong04}, this results in overestimating the UV 
corrected luminosities and consequently to an overestimate of the SFRs. Indeed, when we plot the FIR-to-UV luminosities 
of our galaxies against the spectral slope ($\beta$-slope) we notice that at a fixed $\beta$ all the galaxies have 
systematically lower ratio of total to FIR-to-UV luminosities than the starburst galaxies of \citet{Meurer99}. 
On the other hand, galaxies HCG16c and HCG16d which are infrared luminous ( L$_{\rm IR}>$10$^{11}$L$_{\odot}$), have SFRs which 
are very similar to those derived by the model (10.02 and 1.52M$_{\odot}$yr$^{-1}$ respectively). 

Since, the previous method overestimates the SFR, we can also use the relation of \citet{Buat05}. These authors used the infrared to UV 
luminosity ratio to quantify the dust attenuation in the FUV band of GALEX for a wider sample consisted of more 
quiescent galaxies. The formula they derived to correct for dust extinction is described by the following relation:
\begin{equation}
 A_{\rm FUV}=−0.0333 y^{3} + 0.3522 y^{2} + 1.1960 y + 0.4967
\end{equation}
where y=log(F$_{dust}$/F$_{\rm FUV}$) is the ratio of the infrared to FUV flux densities. To derive these ratios for our 
galaxies we used the infrared luminosities estimated by our SED modelling, as well the observed UV luminosities and estimated the SFR 
using the formula (4) of \citet{Iglesias06} : 
\begin{equation}
{\rm
 log [SFR_{FUV}(M_{\odot}yr^{-1})=0.63\times(log [L_{FUV,corr}(L_{\odot})]-9.51)
}
\end{equation}
where the L$_{FUV,corr}$ is the extinction corrected FUV luminosity (in L$_{\odot}$) and the factor of 0.63 has been introduced 
to correct for the Chabrier IMF  \citep{Chabrier03,daCunha08}. The results are presented 
in Fig.~\ref{fig:sfrs} in green. We notice that the ratios of the SFRs are uniformly distributed around $\sim$1.05 
displaying a scatter of 0.56dex and  also agree with the 24$\mu$m estimates. Given the success of the model in a 
variety of systems \citep[see][]{daCunha08, daCunha10a, daCunha10b} we will base the remaining of the analysis in the model 
derived SFRs.

\section{Results}

\subsection{Evolutionary state of the groups}

In order to study the star formation properties of HCGs groups, \citet{Bitsakis10} separated them into dynamically 
``young'' and  dynamically ``old''. We classified a group as dynamically ``young'' if at least 75\% of its galaxies 
are late-type. Conversely, a group is dynamically ``old'' if more than 25\% of its galaxies are early-type. It is known that the specific 
star formation rate (sSFR), is a tracer of the star formation history of a galaxy, and galaxies in compact groups do 
experience multiple encounters with the various group members. Consequently, we would expect that  a  
young group is more likely to have a larger fraction of late-type galaxies, since its members would not have enough time to experience 
multiple interactions which would trigger star formation, consume the available gas and transform them into early-type systems. 
Furthermore, these late-type galaxies would have not built up much of their stellar mass. They would still have larger amounts of 
gas and dust, as well as higher SFRs and sSFR. On the other hand, if a group is dominated by early-type systems, it would 
be dynamically ``old'', since interactions and possible merging of its members over its history would have led to the 
formation of some of those ellipticals. As a result, the spirals in these groups would have already built some of their stars 
and their sSFR would be lower.  In our present sample, 10 groups (HCG2, 7, 16, 38, 44, 47, 54, 59, 91 \& 100) are classified 
as dynamically ``young'' and the remaining 22 as dynamically ``old'' (see Table~\ref{tab:group}).

We notice that in the majority of the groups (80\%) which are classified as dynamically ``old'', the most massive galaxy 
that dominates is an early-type galaxy displaying a median stellar mass of 1.2$\times$10$^{11}$M$_{\odot}$. In all dynamically 
``young'' groups the dominant galaxy is a late-type, with a median stellar mass of 5.1$\times$10$^{10}$M$_{\odot}$. These 
most massive galaxies of each group, are marked with a star in Table~\ref{tab:sfrs}. Despite the fact that there are just 10 
dynamically ``young" groups in our sample and we could be affected by small number statistics, we note that dynamically ``old" 
groups have on average 4.7 members in each, more than the ``young" groups which only have 4 members.

A different classification of the HCGs into three phases, 1, 2, and 3, based on their HI gas content was proposed by 
\citet{Verdes01}. In the first phase the HI gas is mainly associated with the disks of galaxies. In the second, 40-70\% of the HI is 
found in the disks of the galaxies and the rest has been stripped out, due to tidal stripping, into the intragroup medium. 
Finally, in the third phase are groups with almost all their HI located outside of member galaxies, or in a common 
envelope engulfing most group members. Using this classification \citet{Borthakur10} classified 14 of the groups in our 
sample and the results are presented in Table~\ref{tab:group}. Both our, as well as \citet{Verdes01} classification method, 
are based on how galaxy interactions and merging affect the morphology of the group members, thus they are related to their 
evolution. There is a global agreement between the two methods for 12 of the 14 groups we have in common. We will discuss these 
in more detail in Sect. 4.6.   

\subsection{The physical properties of HCG galaxies}
\begin{figure}
\begin{center}
\resizebox{0.8\hsize}{!}{\includegraphics{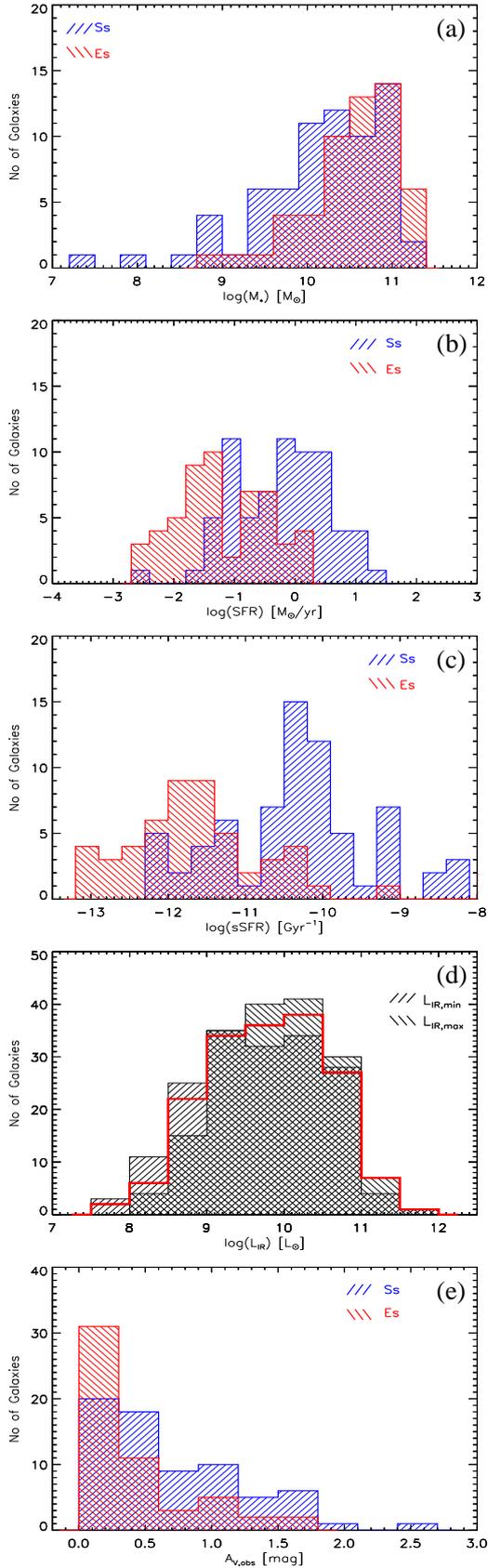}}
\caption{Histograms of several properties of our HCG sample. Early type galaxies 
are marked in red and late type galaxies in blue a) the distributions of the stellar 
masses (M$_{*}$). b) the distributions of the star formation rates (SFR). c)  the 
distributions of specific star formation rates (sSFR). d) The histogram of the infrared 
luminosities, L$_{\rm IR}$, in red, as estimated by our SED modelling. Because of the 
absence of far-IR observations for most of the sample we overplot in hashed areas the 
histograms corresponding to the minimum and maximum  L$_{\rm IR}$ of each galaxy, associated 
to the 16th and 84th percentile range of the best fit respectively. e) A histogram of the 
extinction inferred from the SED fit (A$_{V,obs}$) as measured model-derived atttenuated and unattenuated 
SED (see Section 3.1.3). }
\label{fig:big}
\end{center}
\end{figure}
We will now use the results of the SED modelling, described in the previous section, to examine the physical parameters of the 
galaxies in our sample and compare them with the ones derived for the  comparison samples. The large number of galaxies, a total 
of 135 in 32 groups, is sufficient to produce statistically significant results in order to explore these parameters.

In panel a of Fig.~\ref{fig:big} we present the distributions of the stellar masses separated in early- (E's and S0's) and 
late-type (S's and Irr's) galaxies in groups, marked in red and blue respectively. We can see that the distributions are not too 
different as a two-sided Kolmogorov-Smirnov (KS) suggests that the probability (P$_{\rm KS}$) that the two samples are drawn from 
the different populations is $\sim$0.04. The median stellar masses of the late and early-type galaxies in our sample are 
1.78$^{+0.79}_{-0.55}$$\times$10$^{10}$M$_{\odot}$  and 4.07$^{+0.72}_{-0.60}$$\times$10$^{10}$M$_{\odot}$  respectively. One 
might have expected that the late-type galaxies would have much lower stellar masses compared to the early-types since the latter 
had more time to increase their stellar mass as they converted their gas into star and/or resulted from the merging of late type 
systems . However, as we know tidal interactions play an important role in triggering star formation in galaxies \citep{Struck99}. 
Compact groups have high galaxy density, display  signs of interaction \citep[eg][]{Verdes01} and contain galaxies which are actively
forming young stars. Furthermore, as we showed in the previous section, most of the massive late-type galaxies are found in dynamically 
``old'' groups. We thus believe that several of the HCG late-type galaxies have already increased their stellar mass, due to past tidal 
encounters. We will examine more specifically these galaxies and their properties in the next section. 

In Fig.~\ref{fig:big}b we present the distributions of the star formation rates of early and late-type galaxies in groups. As it was 
expected most of the late-type galaxies have higher SFRs  than the early-type systems,  their median SFR being 
0.60$^{+0.12}_{-0.14}$M$_{\odot}$yr$^{-1}$ compared to  0.05$^{+0.02}_{-0.01}$M$_{\odot}$yr$^{-1}$. The maximum SFR seen in our sample  is  $\sim$26M$_{\odot}$yr$^{-1}$, typical of starbursts in the Luminous Infrared Galaxy (LIRG) range. We should note that there are 8 early-type galaxies with relatively high SFRs ($\sim$1-5M$_{\odot}$/yr) which is more than an order of magnitude higher than what is observed in the rest of the sample. We will 
discuss these galaxies in more detail in Section 4.4.  

In panel c of Fig.~\ref{fig:big} we display the distributions of the sSFRs of our sample. We can see that there are late-type 
galaxies with rather low sSFRs ($\sim$10$^{-12}$yr$^{-1}$). These galaxies, which belong to dynamically 
``old'' groups,  must have had already increased their stellar masses, thus decreasing their  sSFRs, as a consequence of dynamically triggered 
star formation events due to past  interactions. On the other hand, there are early-type galaxies which display rather high sSFRs 
($\sim$10$^{-10}$yr$^{-1}$).  One explanation is that  accretion and merging of gas rich dwarf companions has increased the gas 
content of these galaxies and they are currently forming stars. However, it is also possible that these galaxies are misclassified, 
dust obscured, edge one late-type systems.  In Sections 4.3 and 4.4 we examine in more detail the sSFR distribution as a function of 
the dynamical state of each group, study the mid-IR and optical colors of the galaxies and compare them with our control samples. 

In Fig.~\ref{fig:big}d we present, in red, the distribution of the infrared luminosities (L$_{\rm IR}$) as estimated 
from the \citet{daCunha08} model. Since no far-IR observations were available for most of the galaxies (74\%) of our sample, the 
L$_{\rm IR}$ can not be robustly constrained.  Using for each galaxy the minimum and the maximum values of the probability 
distribution functions as reported by the model, we create two additional histograms corresponding to the low and high 
values respectively. We observe that the overall shape and median value of the distribution does not change substantially. The 
median L$_{\rm IR}$ of the sample is  5.0$^{+3.0}_{-0.9}$$\times$10$^{9}$\,L$_{\odot}$ and most of the HCG galaxies are not IR luminous 
(L$_{\rm IR}\geq$10$^{11}$\,L$_{\odot}$). There are only 7 galaxies, HCG4a, HCG16c, HCG16d, HCG38b, HCG91a, HCG92c and HCG95a,  with 
L$_{\rm IR}>$10$^{11}$L$_{\odot}$. Note that these are all late-type systems, they are detected in the far-IR, and the first 5 are 
found in dynamically ``young'' groups.

As we discussed in  Section 3.1.3 we can use the model-derived attenuated (observed) and unattenuated SEDs to estimate the V-band 
optical depth (A$_{V,obs}$) for each galaxy. Two histograms of these values, for the early and late-type systems, are presented in 
Fig.~\ref{fig:big}d. The median A$_{V,obs}$ is 0.23$\pm$0.17mag and  0.58$\pm$0.36mag for the early and late-type galaxies, 
respectively. We observe that there are 8 ($\sim$14\%) of the early-type galaxies, with extinctions similar to what is seen 
in dusty late-type galaxies ($\geq$1mag). These are HCG4d, HCG55c, HCG56b, HCG56d, HCG56e, HCG71b, HCG79b, HCG100a. We should note  
that these galaxies are the ones forming the tail of early-type galaxies with high SFRs and sSFRs, mentioned above.
The fact that they appear to have larger amounts of dust than what one may expect for early-type galaxies, as well as the 
fact that they display higher star formation activity, further supports the idea that these may indeed be misclassified late-type systems. 

\begin{figure}
\begin{center}
\resizebox{\hsize}{!}{\includegraphics{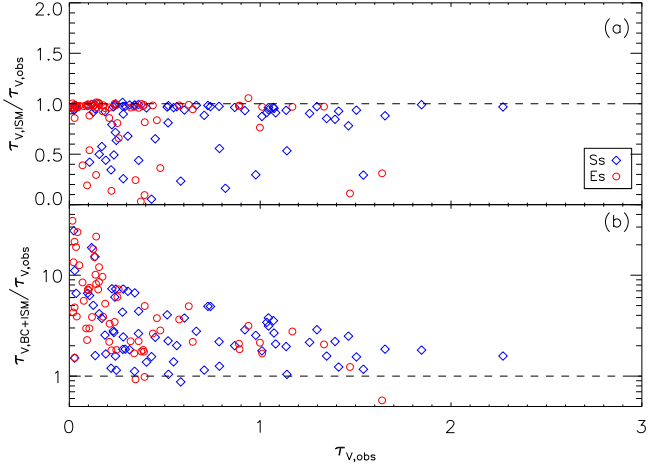}}
\caption{a) Ratio of the intrinsic $\tau_{V,ISM}$ derived by the model over the 
observed (attenuated spectrum) $\tau_{V,obs}$ versus the $\tau_{V,obs}$. The dashed line is 
the one-to-one ratio. The red circles represent the early-type galaxies in HCGs 
and the blue diamonds the late-type ones. b) Ratio of the intrinsic $\tau_{V}$ for stars 
in birth clouds, derived by the model over the observed $\tau_{V,obs}$ versus the $\tau_{V,obs}$, 
using the same notation. We excluded from the plot 6 galaxies (HCG44a, HCG47a, HCG47b, 
HCG56b, HCG59a and HCG68a) since  their $\chi^2$ of model-fit was more than 9.00 and 
the derived $\tau$ fairly uncertain (See Table 3).} 
\label{fig:ratio_Av}
\end{center}
\end{figure}

The model of \citet{daCunha08} also estimates the total effective V-band absorption optical depth of the dust seen by a young 
star ($\tau_{V}=\tau_{V,BC}$+$\tau_{V,ISM}$) inside the stellar birth clouds, and the fraction of the absorption ($\mu$) 
contributed by dust found  in the diffuse ISM  along the line of sight. The emission from young stars is more attenuated than 
that from old stars. In young stars the extinction is dominated by the dust of the surrounding birth clouds with an additional 
component due to the dust in the diffuse ISM traversed by their light. Older stars ($>$ 100Myr), which have dispersed their 
birth clouds, are only affected from the dust found in the diffuse ISM.  Consequently, one would expect that in early-type 
galaxies, where currently no recent star formation is taking place, the optical light would be dominated by the old stars 
and therefore, the representative obscuration would be that of the ambient ISM. In late-type galaxies the total 
extinction would be the contribution of both components, BC and ISM. In that case, each component will contribute differently, 
depending on how much of the optical energy production in the galaxy is due to stars within the birth clouds. We would like 
to emphasize that since both components can  contribute to a different extent at different wavelengths, the representative 
extinction of a galaxy varies also with wavelength, and that is why here we consider only optical light, instead of referring 
to the total energy output of the galaxy. 

Using the output of the model, we can thus estimate the optical depth contributed from 
the diffuse ISM, $\tau_{V,ISM}=\mu\,\tau_{V}$, and the optical depth from the stellar birth clouds, $\tau_{V,BC}=(1-\mu)\,\tau_{V}$. 
In Table~\ref{tab:sfrs} we present the optical depth derived from the attenuated and unattenuated SED, $\tau_{V,obs}$, the $\tau_{V,ISM}$, 
as well as the total optical depth seen by young stars inside stellar birth clouds, $\tau_{V}$=$\tau_{V,BC}$+$\tau_{V,ISM}$. In 
Fig.~\ref{fig:ratio_Av}, we plot the ratios of the intrinsic, model-derived optical depth (ISM and BC+ISM components) versus the observed 
optical depth, $\tau_{V,obs}$, we measured from the SED. We note  that in most galaxies the ratio of  $\tau_{V,ISM}$ over  $\tau_{V,obs}$ is 
very close to unity. For these galaxies, in particularly for those with low extinction values, it is the dust in the ISM 
which determines the overall absorption of the emitted radiation. The contribution of the BC component is a small fraction of 
the overall light.  On the other hand there are 28 galaxies with ratios lower than unity. From these, 18  are classified 
as late-type, half of them belong to dynamically ``young'' groups, and 10 are classified as early-type, 6 of which we believe 
are likely misclassified late-type systems (see Section 4.4 and Appendix).  In the bottom panel of the same figure 
we plot the ratio of total $\tau_{V}$ over the $\tau_{V,obs}$, versus the $\tau_{V,obs}$. We notice that there are galaxies 
with very high $\tau_{V}$ over $\tau_{V,obs}$  values at small $\tau_{V,obs}$, most of which are early-type systems. Even 
though these galaxies have had recent ($<$100Myr) star formation events, this does not dominate their global emission. We do 
note though, that as the $\tau_{V,obs}$ increase, the ratio converges towards unity, implying that the contribution of light 
from young stars in the BC component becomes a considerable fraction of the total emission, making the obscuration 
seen by the newly formed stars more representative of the observed obscuration of the galaxy.

\subsection{HCG late-type galaxies}
\begin{figure}
\begin{center}
\resizebox{\hsize}{!}{\includegraphics{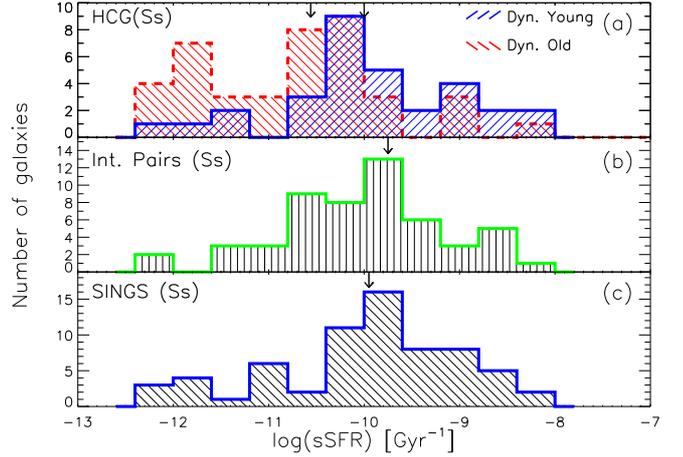}}
\caption{Histograms of the specific star formation, sSFR, of the late-type galaxies our three 
samples, estimated by modelling their SED. The top plot displays in blue  the histogram of the sSFR of 
the 31 late-type galaxies found in dynamically-young, spiral-dominated groups. Over-plotted in red is 
the corresponding histogram  of the 42 galaxies in dynamically-old elliptically dominated groups. 
The middle and bottom  plots present the histograms of the 52 late-type galaxies in the \citet{Smith07a} 
interacting galaxy pairs, as well as the 71 SINGS late-type galaxies. The arrows indicate the median sSFR 
value of each distribution.}
\label{fig:result1}
\end{center}
\end{figure}

It has been well established that interactions can trigger star formation in galaxies \citep[i.e.][]{Struck99}. During an interaction 
between two late type galaxies their atomic gas, typically found in diffuse clouds, collide. Shocks are produced which in turn 
increase the local gas density thus triggering bursts of star formation. The fact that the group environment has played an important 
role in the evolution of its member galaxies, is evident since the fraction of early-type systems in groups is higher that what is 
found in the field. So one would expect that because of their proximity, the late-type galaxies in groups would display different star 
formation properties from the ones in the field. However, in a preliminary analysis of 14 groups, \citet{Bitsakis10} found that 
overall there is no evidence that the SFR and sSFR in late-type galaxies of HCG is different from galaxies in the field or in early-stage 
interacting systems. Interestingly though, when they separated their HCG sample into two sub-samples,  namely the dynamically ``old'' 
and dynamically ``young'' groups (see Sect. 4.1 for the definition), late-type galaxies in dynamically ``old'' HCGs showed lower sSFRs 
than those in dynamically ``young'' groups. This was attributed to the likely larger number of past interactions experienced in the 
dynamically ``old" groups, which would lead to a faster increase of their stellar mass compared to galaxies in young group. The higher 
stellar mass would reduce their current sSFR, even if subsequent  gas accretion would result in star formation activity in them.

We re-examined this issue using our larger sample of 32 groups which contains  73 late-type galaxies and derived their sSFR as well 
as the one of the control samples, using the SED model of \citet{daCunha08}. The results are shown in Fig.~\ref{fig:result1} and 
Table 3. Galaxies in dynamically ``young'' groups have a median sSFR of 8.51$^{+4.07}_{-2.75}\times10^{-11}$yr$^{-1}$, while for 
galaxies in the dynamically ``old'' groups sSFR=2.75$^{+2.03}_{-1.16}\times10^{-11}$yr$^{-1}$. Similarly, galaxies in interacting 
pairs have a sSFR=11.20$^{+3.67}_{-2.70}\times$10$^{-11}$yr$^{-1}$ and in field galaxies sSFR=15.30$^{+5.65}_{-4.29}\times10^{-11}$
yr$^{-1}$. An analysis using two sided KS test indicates that there is no statistical difference between the samples of late-type 
galaxies in dynamically ``young'' HCGs and those of the SINGs and interacting pair samples (P$_{\rm KS}$$>$0.80). However, 
the same KS test reveals that the late-type galaxies in  dynamically ``old''  groups, having a median sSFR which is more than three times  
lower, can not be drawn from same parent distribution as the other three samples (P$_{\rm KS}$$\sim$10$^{-3}$). Investigating in more 
detail the reason for this disparity, we find that it cannot be attributed to depressed SFR but instead it is due to a substantially more 
massive stellar content ($\sim$3$\times$10$^{10}$M$_{\odot}$), similar to what is found early-type systems. This confirms the 
results and interpretation of \citet{Bitsakis10} who relied on semi-empirical estimates of the sSFR on a smaller galaxy sample. Based on 
those findings one would also expect that due to their higher stellar mass, the late-type type galaxies in dynamically ``old'' 
groups should have redder UV and optical colors than late-type galaxies in the field. We examine this in Section 4.5.

\subsection{HCG early-type galaxies}
\begin{figure}
\begin{center}
\resizebox{\hsize}{!}{\includegraphics{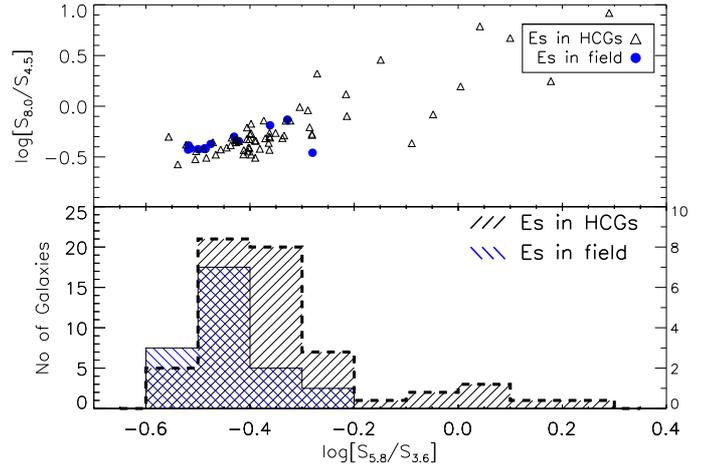}}
\caption{Top panel: IRAC color-color plot of the early-type galaxies in HCGs (black open 
triangles) and a sample of field ellipticals (blue filled circles). Bottom panel: 
The histograms of the log[S$_{5.8}$/S$_{3.6}$] color of the two samples. The  
early-type galaxies in the groups are marked with a black line (left axis) while the field elliptical 
are presented in blue  (right axis).}
\label{fig:result2}
\end{center}
\end{figure}

We showed in the previous section that the group environment  affects the star formation history of the late-type 
members increasing their stellar mass  in the process of transforming them into early-type systems. However, is there 
any evidence indicating that the early-type galaxies in groups are distinctly different from similar galaxies found in the 
field? Based on our SED modelling, we have already showed in Fig.~\ref{fig:big}, that nearly 15\% of all early-type galaxies 
have SFRs and dust extinction similar to what is seen in late-type systems. We suggested that there are two possible explanations 
for this. One was that they were simply misclassified as Es or S0s, while in fact they are edge-on late-type systems. 
The second was that these galaxies even if they are early-type, based on their optical morphology, they have experienced minor 
merging with gas/dust rich dwarf companions in the groups, which has increased their gas content, star formation activity, and 
dust extinction. 

According to \citet{Hickson82} the classification of the HCG galaxies was performed using their morphological features, 
optical colors, and sharpness of the edge of the image. For all groups of our sample though, we also have Spitzer mid-IR images 
which provide additional information on their properties. It is well known that the mid-IR spectra of the 
star forming galaxies in addition, to continuum emission due to warm dust, are also filled with a series of broad emission features 
between 3 to 18$\mu$m, which can contribute up to 20\% of their total IR luminosity \citep{Dale05,Smith07b}. These features are the 
vibrational modes of Polycyclic Aromatic Hydrocarbons (PAHs), which absorb UV photons from newly born stars and re-emit them 
in the mid-IR. Typically only late-type galaxies with active star formation and dust emit strongly in these bands. Some PAHs 
have also been detected in some early-type systems, but they display peculiar PAH-band ratios \citep{Bressan06,Kaneda08,Panuzzo11}. To 
probe the properties of the early-type galaxies in the groups, we examine their mid-IR colors and compare them with isolated 
ellipticals in the field which are not expected to have PAH emission. In Fig.~\ref{fig:result2} we use our Spitzer/IRAC 
photometry and plot the 8.0 to 4.5 $\mu$m flux density ratio as a function of the 5.8 to 3.5$\mu$m ratio for the early-type galaxies of 
our HCG sample, as well as the control sample of field ellipticals. Weak PAHs would result in low values of both ratios 
since the 6.2 and 7.7$\mu$m features, which contribute $\sim$50\% of the total PAH emission,  are sampled by  the 5.8 and 8.0$\mu$m IRAC 
filters \citep{Smith07b}. Indeed, we observe that most of the early-type galaxies in groups, as well as in the field are concentrated in 
the lower left part of 
the color-color plot of Fig.~\ref{fig:result2}. These galaxies are very close to the (-0.4, -0.5) locus of pure stellar 
photospheric emission where the flux density in the mid-IR scales with $\lambda^{-2}$. Consequently, they are expected to 
have very weak dust and PAH emission. On the other hand, the upper right quadrant of the figure should be populated by 
galaxies with strong PAH features and possibly some hot dust contribution, as a result of intense star formation and/or 
AGN activity. As expected, no field ellipticals are seen in this quadrant. However, there is a ``tail'' of 10 early-type HCG 
galaxies ($\sim$16\% of the total) extending to this part of the plot (for log[f$_{5.8}$/f$_{3.6}$]$>$-0.25). Examining the 5.8 to 3.5$\mu$m IRAC color distributions of HCG and field  with a KS test, we find that they are different (P$_{\rm KS}$$\sim$0.005). The early-type galaxies with red IRAC colors are: HCG4d, HCG40f, HCG55c, HCG56b, 
HCG56d, HCG56e, HCG68a, HCG71b, HCG79b \& HCG100a. We note that HCG40f and HCG68a are at the left edge of  the color selection 
and it is possible that contamination from a red nearby companion is affecting their mir-IR fluxes. All these galaxies are 
examined in more detail in the Appendix, where we display their SEDs, Ks-band contour plots and ``true color" composites based 
on their Spitzer/IRAC mid-IR images. Based on this analysis and previous suggestions, we propose that 7 of them, HCG4d, HCG55c, HCG56d, HCG56e, HCG71b, 
HCG79b \& HCG100a  are likely dust obscured late type systems. Furthermore, if  HCGd and HCG71b are indeed  late-type galaxies, then their 
groups must be reclassified as dynamically ``young'' (instead of ``old"). In the remaining of the paper we will considered them as such (see Table 6 and Table A.1).

\subsection{Bimodality in HCG galaxy colors}

\begin{figure}
\begin{center}
\resizebox{\hsize}{!}{\includegraphics{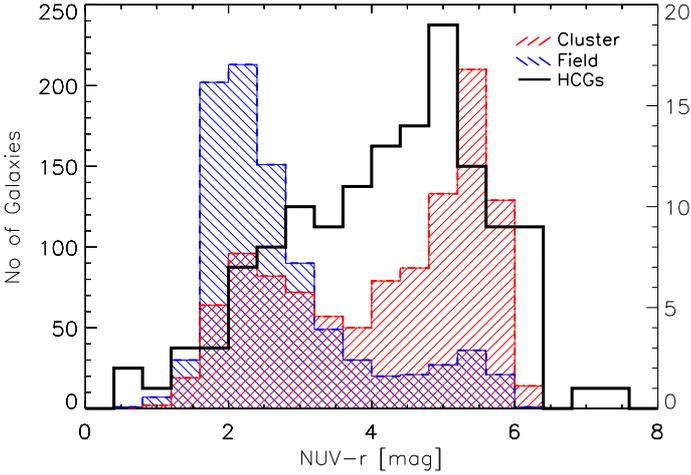}}
\caption{NUV-r histogram of our HCG galaxy sample shown in black (right Y-axis).
The histograms of the of field, in blue, and cluster galaxies, in red, from the
\citet{Haines08} control sample are also presented (left Y-axis).}
\label{fig:result3-1}
\end{center}
\end{figure}

In the previous sections we have examined several of the physical parameters of our HCG sample as derived by the SED modelling. In order 
to examine more thoroughly the evidence of evolution due to the group environment, we will now compare all our results  based on the UV, 
optical, and infrared imaging as well as the available spectroscopic diagnostics.

A number of studies of galaxies have revealed a bimodality in the UV-optical colors of galaxies in field and in clusters
\citep[e.g.][]{Wyder07,Haines08}. This type of bimodality in the colors of the late- and early-type galaxies is real and 
directly connected to the original galaxy classification scheme by Hubble. It appears quite strongly in the NUV-r color
distribution of a sample and consists of two peaks, called the ``red sequence'' and the ``blue cloud'', and a minimum in
between them identified as the ``green valley'' \citep[see][]{Strateva01}. \citet{Wyder07} examined a sample of over 18,000
galaxies observed with GALEX and SDSS and showed that a NUV-r versus M$_{r}$ color magnitude diagram can be used to separate
them into (i) passively evolving, (ii) star-forming and (iii) AGN components. Passively evolving galaxies are well
confined to the ``red sequence'' (NUV-r$>$5), with few showing blue UV-optical colors,  while all blue galaxies, with NUV-r$<$3,
are spectroscopically classified as star forming \citep[see also][]{Haines08}. AGN seem to dominate the area of the green valley''
(3$<$NUV-r$<$5). The small fraction of galaxies in the ``green valley'' can be understood from the fact that galaxies do
not spend so much time at these colors. Using  Starburst99  \citep[see][]{Leitherer99} we simulated the evolution of a stellar population 
typical of a spiral galaxy assuming solar metallicity and either a continuous star formation (of 1M$_{\odot}$yr$^{-1}$) or a single starburst 
(producing total stellar mass of 10$^{6}$ M$_{\odot}$). Depending on the assumptions, we find that the colors of the population place it 
within the ``blue cloud'' just for $\sim$5-7Myr and after spending  only $\sim$1-5Myr transiting through the ``green valley'' it remains 
in the ``red sequence'' for the rest of its lifetime. Thus this color bimodality emerges from the nature of the galaxies as they evolve 
through time.


\begin{figure}
\begin{center}
\resizebox{\hsize}{!}{\includegraphics{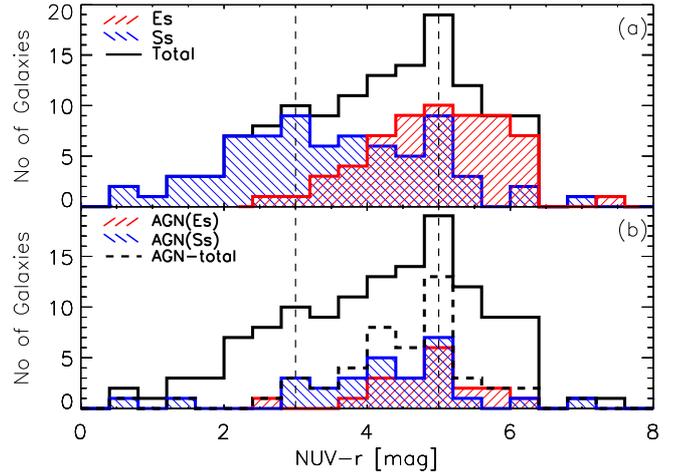}}
\caption{a) NUV-r histogram of the whole HCG galaxy sample is shown in black. The
corresponding histograms of the early- and late-type galaxies of the sample, classified
according to their optical morphology, are shown with the red and blue shaded areas. b)
The NUV-r distribution of the galaxies hosting an AGN is indicated with the dashed dark
line. As above, the early- and late-type hosts are indicated with the red and blue
histograms. The region of 3$<$NUV-r$<$5, identified as ``green valley", is marked with
the vertical dashed lines.}
\label{fig:result3-2}
\end{center}
\end{figure}

In Fig.~\ref{fig:result3-1} we plot the NUV-r colors of our HCG sample, as a black solid line, and compare them with the
field and cluster galaxies of the \citet{Haines08} sample, indicated in blue and red respectively. The three regions of the color
space discussed by \citet{Wyder07} are clearly visible. On the left, near NUV-r$\sim$2,  we identify the ``blue cloud'' where the
field galaxy distribution peaks. Nearly 58\% of the total number of field galaxies is found within 0.5mags of this color. Few
of the galaxies in clusters (15\%) are also seen in this area. To the right, near NUV-r$\sim$5.5, we observe the  ``red sequence''
where most (51\%) of  the cluster galaxies are concentrated. Only 16\% of the field galaxies are found in the ``red sequence''.
The ``green valley'' is between the two peaks and is mostly (34\%) populated by cluster galaxies \citep{Haines08}. When we
examine the HCG NUV-r color distribution, we note a clear difference. Only  17\% of their galaxies  are in the ``blue cloud'', a fraction similar to
what is found for galaxies in clusters. However, the majority (73\%) of the HCG galaxies are concentrated in the ``green valley''
and ``red sequence'' areas. A likely explanation would be that most of the HCG galaxies are already passively evolving or they are
in the process of moving from the star-forming region to the ``red sequence''. This is consistent with the fact that compact groups have
a higher fraction of elliptical galaxies than the field. Another possibility is that some of these galaxies are in fact late-type
systems which have redder colors and hence move to the right on the plot, because of dust extinction and/or due to the presence of
a substantial older stellar population (post starburst galaxies). This is expected from our earlier findings regarding the
miss-classification of a fraction of the early-type galaxies in the HCG and the lower sSRF in dynamically ``old'' groups.

To study in more detail how the NUV-r color distribution of the HCG galaxies is affected by their morphology and nuclear
activity, we present in Fig.~\ref{fig:result3-2} the corresponding histograms based on their optical morphologies, as well as
their nuclear spectral classification. We note that only 36\% of the late-type galaxies, indicated in blue, are found within
the ``blue cloud''. However the majority of the HCG galaxies, almost 60\%, are located in the ``green valley'' and in the
``red sequence'' ($\sim$13\%). In  Fig.~\ref{fig:result3-2}b we present the same plot as at the top, but only for the galaxies of
our sample who have an optically identified AGN. The whole distribution is shown with the black dashed line, while the distributions of
early- and late-type AGN hosts are shown in red and blue. We performed a two sided KS test between the distribution of galaxies hosting an
AGN and the total distribution of the HCG galaxies and we conclude that there is no statistical difference in their NUV-r colors
(P$_{\rm KS}$$\sim$0.25). This suggests that the presence of an AGN does not affect substantially the UV-optical color of the galaxies in
Hickson Compact Groups.

As we described earlier, more than 40\% of the galaxies in our sample host an AGN into their nucleus. This, in addition with the fact that 
the code of \citet{daCunha08} doe not include the contribution of an AGN to the SED of a galaxy, could bias some of our 
results, an in particular the SFR, sSFR and the L$_{\rm IR}$. Using the IRAC color-color AGN diagnostics introduced by \citet{Stern05} we investigated the 
influence of the AGN to the mid-IR SED of the galaxies in our sample. Only  three systems, HCG6b, HCG56b and HCG92c, have mid-IR SEDs which are consistent with a strong power-law continuum emission indicative of a strong AGN in their nucleus. For all remaining galaxies the optically identified AGN does not dominate their SED and thus their physical parameters estimated by our model are considered reliable.

\begin{figure}
\begin{center}
\resizebox{\hsize}{!}{\includegraphics{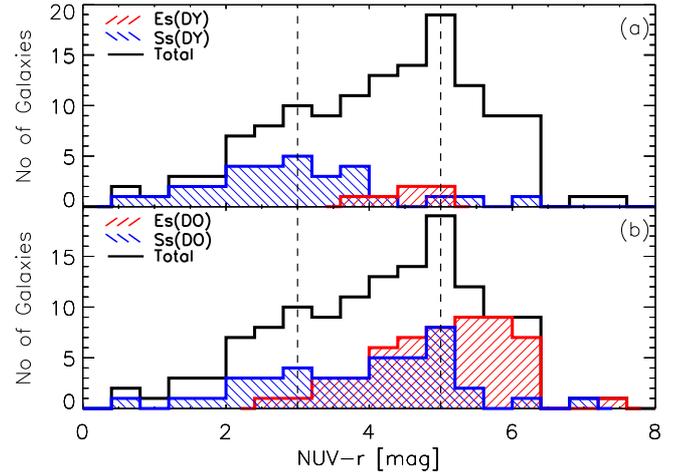}}
\caption{a) NUV-r histogram of our HCG galaxy sample shown in black solid line. The
corresponding histograms of the early- and late-type galaxies found in dynamically
``young'' groups are shown with the red and blue shaded areas respectively. b) Same as
in a), but for the galaxies in dynamically ``old'' groups.  The region of 3$<$NUV-r$<$5,
identified as ``green valley", is marked with the vertical dashed lines.}
\label{fig:result3-3}
\end{center}
\end{figure}

In order to explore the color variations as a function of the evolution state of the groups, we plot in Fig.~\ref{fig:result3-3} the histograms of the early-
and late-type galaxies found in the dynamically ``young''  and ``old''  groups respectively. Observing the top panel we find that
almost 60\% the late-type galaxies in dynamically ``young'' groups are located within the ``blue cloud'' and 43\% of them, for which
nuclear spectra were available, host an AGN in their nucleus. There are also three outlier galaxies (HCG16b, HCG44a and HCG59a)
which have red NUV-r colors ($>$5mag). It is possible, that these systems have built up their stellar mass in the past and their
UV/optical colors are currently dominated by emission from old stars. In addition, past tidal interactions probably stripped some
of their gas in the intragroup medium decreasing the fuel necessary for current star formation. In  dynamically ``old'' groups the
late-type galaxies are redder and as we can see in Fig.~\ref{fig:result3-3}b most of them ($>$63\%)  are located within the
``green valley''. As in dynamically ``young'' groups there are also four galaxies (HCG22b, HCG40d, HCG68c and HCG71a) in these
groups which are found in the ``red sequence''.

Overall 45\% of the early-type galaxies, shown in red in Fig.~\ref{fig:result3-2}, are located within the ``red sequence''
and the rest are along the ``green valley''. As we mentioned earlier, these systems are expected to be passively evolving and thus, their 
colors should be dominated by old stellar populations. We find that this is the case for most of them, as only two are found in the ``blue cloud''. We believe that
these galaxies are misclassified late-type galaxies (see Sect. 4.4) and we examine them in more detail in the Appendix. From Fig. ~\ref{fig:result3-3}a we
see that all early-type galaxies in the dynamically ``young'' groups are in the in the ``green valley''. One could suggest that they
were already early-type before reaching the group and the interactions and/or merging of dwarf companions in the group environment triggered 
the star formation activity and provide extra gas. However, the most possible explanation is that these are group galaxies which are currently
migrating from the ``blue cloud'' to the ``red sequence''. In that case, the morphological transformation should accompany the color
transformation. Indeed, the galaxies located in bluer colors (around NUV-r$\sim$4) are lenticulars, while the ones at
NUV-r$\sim$5 are ellipticals. Similarly, at the bottom panel of Fig.~\ref{fig:result3-3} we see that more than half of the early-type galaxies
in dynamically ``old'' groups are located within the ``green valley''. These galaxies are mostly ($>$70\%) S0/SB0's. However, there is a large
fraction ($\sim$25\%) of elliptical galaxies which also have the same colors. We suggest that even though the morphology of a galaxy typically
determines its colors, there are several of the group ellipticals which move back to bluer colors due to interactions and/or merging in the
compact group environment.

\begin{figure}
\begin{center}
\resizebox{\hsize}{!}{\includegraphics{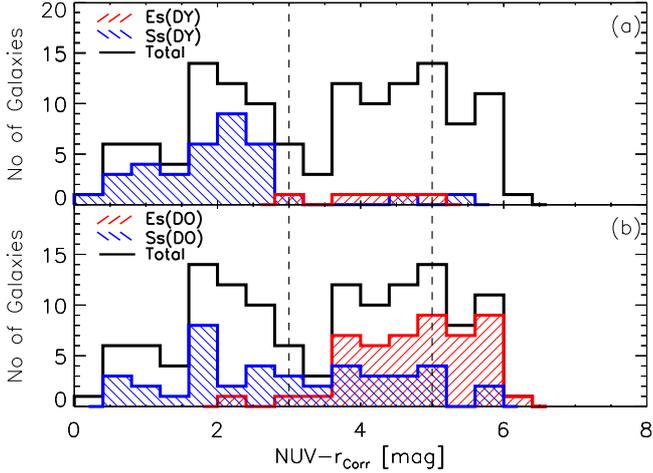}}
\caption{A histogram of the NUV-r colors of the HCG sample after correcting for dust 
extinction, in shown with the black solid line. The separation in two panels for ``young" 
and ``old" groups as well as the symbols follow the notation of Fig.~\ref{fig:result3-3}
}
\label{fig:result3-4}
\end{center}
\end{figure}

As discussed before, we know that the UV-optical colors of a galaxy, are not only affected by its star formation history, but also
from dust extinction. Thus, it would be useful to understand how much of the observed variation
in colors is due to intrinsic extinction in the galaxies. Especially for the late-type galaxies which are located within the ``green 
valley'', we have to examine if their colors are affected more from  dust  or because they contain a substantial old stellar population. Our
SED model allow us to estimate and correct for dust attenuation, thus we plot in Fig.~\ref{fig:result3-4} the distributions of the corrected
NUV-r colors of the early- and late-type galaxies in dynamically ``young'' (top panel) and dynamically ``old'' groups (bottom panel).
We notice in panel a that almost all late-type galaxies in ``young'' groups move back to the ``blue cloud'', suggesting that it is the 
dust which plays a dominant role in their apparent colors. On the other hand, 40\% of the late-type galaxies in dynamically ``old'' groups 
still remain in the ``green valley'' after the extinction correction. It seems that in those it is the old red stars that dominate their 
UV/optical colors. This also agrees with the results of Section 4.3 where these galaxies display low sSFRs. Moreover, one can
notice the large fraction of early-type galaxies which after the extinction correction move to the ``green valley'' ($\sim$50\% of the
galaxies at these colors). These galaxies display high sSFRS ($\sim$0.21$\times$10$^{-11}$yr$^{-1}$), so their bluer colors
are possibly due higher levels of current star formation. We should note that even though for most galaxies the observed optical extinction 
derived from the SED model (A$_{V}$) is not very high (see Fig. 4e and Table 3), due to the steepness of the extinction curve this translates 
to over 2mags of correction in the NUV.

\begin{figure}
\begin{center}
\resizebox{\hsize}{!}{\includegraphics{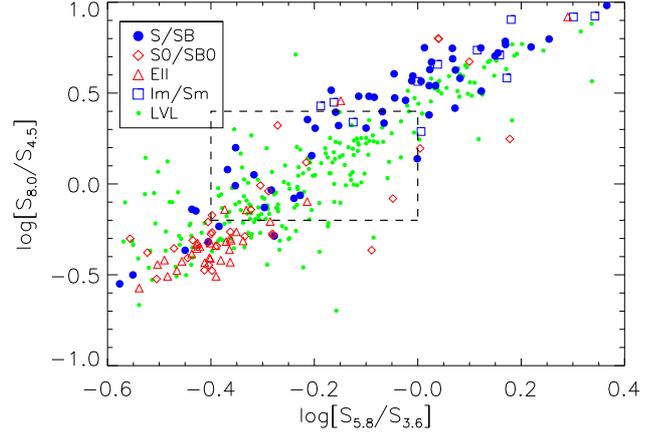}}
\caption{The IRAC color-colors plot, originally presented in Fig. 7. Blue circles indicate the spiral galaxies of our HCG sample,
blue squares are the irregulars, red triangles are the elliptical galaxies and red diamonds are the lenticular ones. Green dots
indicate the LVL galaxies. According to \citet{Johnson07} the ``gap'' is located in the area enclosed by the dashed lines.}
\label{fig:irac-color}
\end{center}
\end{figure}

Since dust appears to affect the UV colors of HCG galaxies, one could suggest the use of their mid-IR colors, because they trace
the light which was originally absorbed by the dust grains in the UV-optical. In Fig.~\ref{fig:irac-color} we present
the IRAC color-color diagram, also shown in \citet{Bitsakis10}, using our new, larger sample. The physical meaning of this plot
was described extensively in Fig.~\ref{fig:result2}. We notice that most of the late-type galaxies are located in the upper right quadrant
of the plot, while most of the early-types are in the lower left. The 10 early-type galaxies which display red mid-IR colors are the ones
we mentioned in Section 4.3. We observe that between the colors -0.1$<$log[f$_{8.0}$/f$_{4.5}$]$<$0.3 and -0.25$<$log[f$_{5.8}$/f$_{3.6}$]
$<$-0.10 there is a lower density of galaxies. \citet{Johnson07} and \citet{Tzanavaris10} proposed that this ``gap''  is related to an 
accelerated migration of the galaxies from star forming to quiescent. When we compare this result to Fig.~\ref{fig:result3-4} we notice 
that the galaxies in the upper right quadrant of the IRAC color-color plot are the ones with NUV-r$<$2.5, while galaxies which are 
located in the lower left portion of the figure have NUV-r$>$3.5. We finally notice that the lower galaxy density area appears in both 
figures and actually separates the star forming (``blue cloud'') from the passive evolving (``green valley'' and ``red sequence'') 
galaxies. Therefore we suggest that the color bimodality observed in the extinction corrected UV-optical colors, is also observed in the 
mid-IR and possibly emerges from the same physical properties of the galaxies.

How significant though is this ``gap"? \citet{Walker09} compared their sample of 12 HCGs with several control samples including the SINGS 
and interacting pair sample we used, as well as galaxies from the Coma cluster, and the Local Legacy Volume sample \citep[LVL][]{Dale09}. 
They concluded that the mid-IR color distribution of HCG galaxies is not seen in the field but it is seen in galaxies of the Coma infall 
region. However, in their analysis they combined the SINGS and LVL galaxies as a single local galaxy sample. As we mentioned though in 
Section 2.6.1, the SINGS sample was selected to explore the mid-IR properties of various galaxy types seen in the local universe, but 
it is not representative of a flux or volume limited population \citep{Kennicutt03}. This selection biases the statistics of the colors 
of the sample. In contrast the LVL sample is volume limited, since it contains all known galaxies inside a sub-volume of 3.5 Mpc and an 
unbiased sample of spiral and irregular galaxies within the larger, more representative, 11 Mpc volume. The mid-IR 
colors of the LVL sample are plotted in Fig.~\ref{fig:irac-color} as green points.  Performing a two sided KS-test between the distributions 
of the IRAC colors (log[f$_{8.0}$/f$_{4.5}$] and log[f$_{5.8}$/f$_{3.6}$]) of HCG and LVL samples, we find that the distributions are not 
significantly different displaying P$_{KS}=$0.035 and P$_{KS}=$0.028, respectively. This suggests that there is no strong evidence that an 
accelerated evolution of galaxies in HCG is responsible for a significant fast change in their global mid-IR colors.

\subsection{Dynamical properties of HCG galaxies}

As we mentioned in Sect. 4.1 we classified the dynamical state of our sample depending on the fraction of elliptical galaxies they 
contain. Our method is related to their evolution since it is known that galaxy interactions and merging, transform galaxies from 
late-type systems to ellipticals. Thus, groups with excess of early-type members, are likely to be older and more compact. In 
order to investigate this we have estimated the projected distance of each galaxy from the other members of the same  group members, 
as well as the virial radius of each galaxy. To calculate the projected distance between the group members we used the great circle 
distance relation. Then, using the relation described in  \citet{Park09}, we estimated the virial radius with the formula: 

\begin{equation}
 r_{vir}(\rm{Mpc})=\left( \frac{3\gamma L \Omega_{m}}{800\pi \rho_{m}}\right) ^{1/3}
\end{equation}
where L=10$^{-0.4(\rm M_{r}+20.00)}$ is the r-band luminosity in L$_{\odot}$ and M$_{r}$ is the absolute r-band magnitude of the galaxy, 
$\gamma$ is the mass to light ratio (1 or 2 for  late- and early-type galaxies, respectively), and $\rho_{m}$=0.0223$h^{3}$Mpc$^{-3}$ 
is the mean density of the universe, with $h$=0.72 and $\Omega_{m}$=0.27. Keeping in mind that compact groups may not be fully virialized 
dynamical systems, we present our results in Table~\ref{tab:dynnei}. We note that indeed, in dynamically ``old'' groups the member galaxies 
are statistically closer to their nearest neighbors, having a mean projected distance of 26kpc compared to  37kpc for the dynamically 
``young'' groups. 

\begin{figure}
\begin{center}
\resizebox{\hsize}{!}{\includegraphics{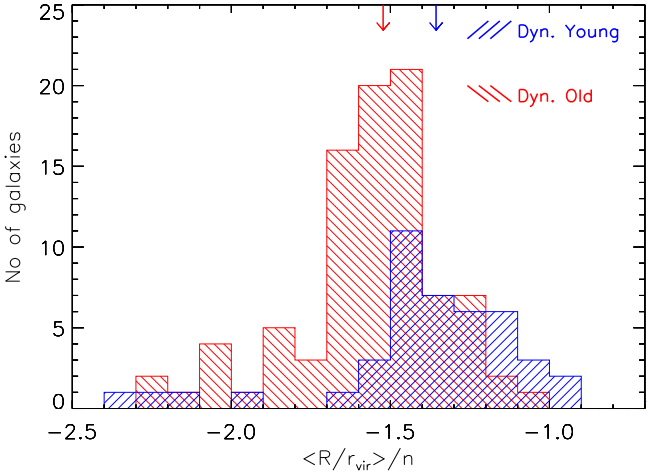}}
\caption{Histograms of the ``strength'' of the dynamical interaction felt by each group 
galaxy ($<$R/r$_{vir,nei}$$>$/n). As usual, we separate our sample into dynamically 
``young'' (blue) and dynamically ``old'' (red) groups. The arrows indicate the median 
values of the two distributions.}
\label{fig:result6-1}
\end{center}
\end{figure}

However, galaxies in groups are interacting not only with their nearest neighbors, but also with all other group members since
in all cases the virial radii of the galaxies are substantially larger than the linear size of their group (see Table~\ref{tab:dynnei}). 
To better quantify this effect we have followed the approach of \citet{Hwang10}, introducing the parameter of the ``strength'' 
of the interaction ($<$R/r$_{vir,nei}$$>$/n). This is estimated by averaging the ratios of the projected distances over the virial 
radii for all the neighbors of each galaxy and divide them by the total number of neighbors ({\it n}). This quantity provides an empirical 
measure of the interaction strength each galaxy experiences from its companions. The smaller this parameter is, the stronger the 
dynamical influence of the companions. 

 In Fig. 13, we plot the distributions of the “strengths” according to the dynamical state of each group. Since we have SDSS data for only 
74 galaxies and equation (6) relies on the SDSS r-band magnitude to compute the virial radius, we used the SED model fits to estimate the 
synthetic r-band magnitudes for the remaining of the sample.
We observe that the ``strength'' parameter of dynamically ``old'' groups is indeed smaller than in dynamically ``young'' groups, suggesting 
that in the former the gravitational effects are stronger. A KS test between the two distributions confirms 
that the two distributions are significantly different (P$_{\rm KS}$$\sim$10$^{-5}$). Moreover, if we compare our results 
with the findings of \citet{Hwang10} we see that the distances between the galaxies in groups and those of the centers of rich clusters 
are similar (both have R$<$0.5r$_{vir}$). We can also estimate the velocity dispersion, $\sigma_{r}$, of each galaxy from the most 
massive-central galaxy of its group, using the published velocities from \citealt{Hickson92}. Galaxies in dynamically ``old'' 
groups display mean velocity dispersion of 408$\pm$50km s$^{-1}$, which is the half of what is observed in the Virgo cluster, in 
contrast with those in dynamically ``young'' groups for which is about 132$\pm$39km s$^{-1}$. All these facts lead us to the 
conclusion that dynamically ``old'' groups, as their name implies, are older, denser and more evolved systems, since their properties 
are intermediate between those of rich clusters and dynamically ``young''  groups.

To further examine the dynamical properties of these groups, we can calculate their dynamical masses using the projected mass 
estimator from \citet{Bahcall81}: 
\begin{equation}
\rm{
M_{dyn}=\frac{24}{\pi GN}\sum\limits_{i=1}^N u_{i}^{2}R_{i}, 
}
\end{equation}

where u$_{i}$ is the difference in the recessional velocity between the group member {\it i} and the central massive object (see 
Table~\ref{tab:sfrs} where the most massive objects in each group are marked with a star), R$_{i}$ is the projected distance 
from the central object, and N is the number of group members. This dynamical mass can be compared to the various components 
of the baryonic mass of the groups. The stellar mass is derived from our SED modelling while the total atomic hydrogen mass, 
M$_{\rm HI}$,  is available for 29 of the groups from  \citet{Verdes01}. Unfortunately, the mass of the molecular hydrogen,  
M$_{\rm H_2}$,  is only available for 12 groups \citep{Martinez-Badenes11}.  We note that even though the integrated 
masses of the molecular and atomic hydrogen of the groups are often similar, on average they are both  $\sim$25 times less 
massive that the stellar component.  These results are summarized in Table~\ref{tab:group}. 

\begin{figure}
\begin{center}
\resizebox{\hsize}{!}{\includegraphics{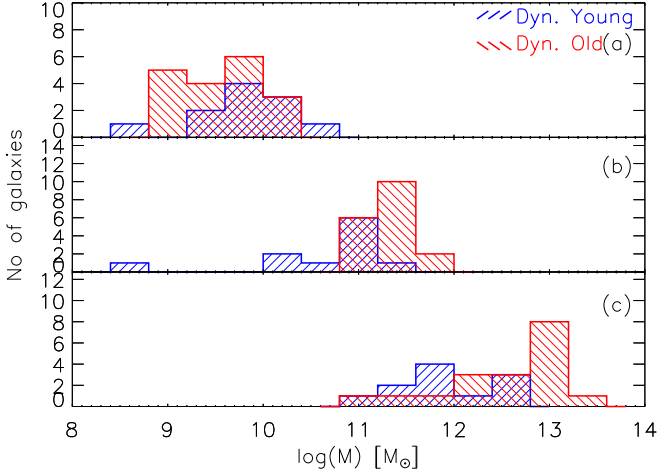}}
\caption{a) Distributions of the total HI masses in the dynamically ``young'' (in blue) and the 
dynamically ``old'' (in red) groups, of the total stellar masses (panel b) and the 
total dynamical masses (panel c).}
\label{fig:result6-2}
\end{center}
\end{figure}

In Fig.~\ref{fig:result6-2} we separate our 32 groups in the two subsamples of dynamically ``young" and ``old"  groups and present 
the corresponding  histograms of the  distributions of their atomic hydrogen, stellar, and dynamical masses.

We observe that the stellar mass distribution of dynamically ``old'' groups, shown in Fig.~\ref{fig:result6-2}b  is markedly 
different from that of dynamically ``young'' groups. A KS test confirms this, as the probability that the two are drawn from 
the same parent distribution is P$_{\rm KS}$$\sim$0.009. The median stellar mass of the dynamically ``old'' groups is  $\sim$1.95
$\times$10$^{11}$M$_{\odot}$, more than double of what is seen in the  ``young'' group subsample ($\sim$8.90$\times$ 
10$^{10}$M$_{\odot}$). This is understood on the basis of our classification scheme; dynamically ``old'' groups do have more elliptical 
galaxies which are typically more massive than spirals. Furthermore, from the analysis of Section 4.3, we have shown that the 
late-type galaxies of these groups also have redder optical colors. 

Surprisingly though, in  Fig.~\ref{fig:result6-2}a we note that the HI mass distribution is fairly similar in both subsamples 
(P$_{\rm KS}$$\sim$0.26), with a median value of HI mass $\sim$3.2$\times$10$^{9}$M$_{\odot}$. Assuming that dynamically ``old'' 
groups are older structures, with more early-type members which have already converted during a merging process a fraction of 
their gas into stars, it is not clear if what we see is a real ``conspiracy'' given the difference in star formation properties 
and relative amount of old stellar populations of the two subsamples. 
We know that a fraction of the HI gas is expelled from the galaxies to the intragroup medium  in the form of plumes and long filaments 
as a group evolves progressively transforming the morphology of its members  from late- to early-type \citep{Verdes01}. However, the 
lifetime of these structures before they either change phase, as they interact with the hot intragroup gas, or are accreted back 
to the group members is not well constrained, since only a dozen of HCGs have high spatial resolution maps of their gas component. 

Examining the distributions of the dynamical masses of the two subsamples in panel c of the same figure, we find that, as was the 
case with the stellar mass, these are also different (P$_{KS}$$\sim$0.008).  The dynamically ``old'' groups display a median dynamical 
mass of 6.46$\times$10$^{12}$ M$_{\odot}$, nearly an order of magnitude higher from what is seen in the ``young" ones 
(4.77$\times$10$^{11}$M$_{\odot}$). Note, that the dynamical masses do not simply scale with the stellar mass of the subsamples, which 
only differ by a factor of $\sim$2. Furthermore, there is no indication of a variation in the ratio between the dark matter, as probed by 
the kinematics, and the baryonic matter in groups, or that there is a systematic error in the estimates of the dynamical mass due to one 
subsample being less virialized than the other. As a result, one could attribute the deficit in the observed baryons in the dynamically 
``old'', to a component of hot intragroup gas, since there is clear evidence that groups with more early type systems have stronger X-ray 
emission \citep{Ponman96,Mulchaey03}. 

In Table~\ref{tab:group} we also present our classification of the 32 groups, based on the fraction of the early-type 
members, as well as the classification  based on their HI morphology proposed by \citet{Verdes01} for 13 groups where this was 
available. Despite the small number statistics, we note that there is an overall agreement. Groups which we consider as dynamically 
old are in Phase 3 or 2 according to their HI content, while groups which we consider dynamically ``young'' are in Phase 1 or 2. 
There are two exceptions in this. HCG79, the so called ``Seyferts' Sextet'' was  classified by us as dynamically ``old" but it is 
Phase 1 based on the distribution of its HI gas by \citep{Borthakur10}. Inspecting it closely we see that our definition is likely 
marginal as the system does not have an elliptical galaxy and its 2 early type members are peculiar S0s. The other case is HCG44, 
which was classified as Phase 3 by \citet{Borthakur10}. The group consists of 3 late type galaxies and NGC\,3193, an E2 which 
\citet{Aguerri06} argue that may be an interloper. Close inspection of the HI map of the group shown in Fig. 2 of \citep{Williams91} 
reveals that nearly all gas emission is associated with the 3 late type systems, and only  a small fraction is seen outside the 
galaxy disks. We would thus argue that this system is more likely in Phase 1 or 2, which would be consistent with our classification 
as dynamically ``young". Therefore, as expected, a change in the galaxy morphology also carries a change in the distribution of the 
atomic hydrogen within the group, and viceversa.

\subsection{Diffuse Cold Dust in the Intragroup Medium}

We mentioned earlier that using the model of \citet{daCunha08} we were able to derive several of the physical properties of the 
galaxies in our sample. Even though a number of parameters such as stellar mass and SFR, can be estimated with good accuracy,  
the lack of deep high spatial resolution wide field far-IR imaging, makes it very challenging to constrain the L$_{\rm IR}$ 
and accurately estimate the amount of cold dust in each galaxy. 

\begin{figure}
\begin{center}
\resizebox{\hsize}{!}{\includegraphics{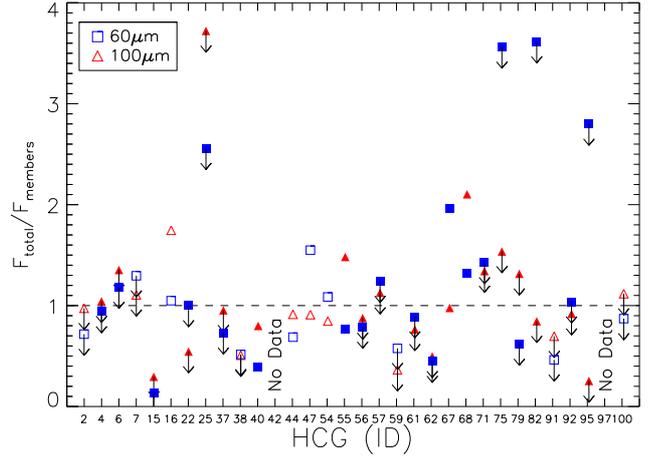}}
\caption{Plot of the ratio of the observed IRAS 60 and 100 $\mu$m fluxes of each 
group divided by the corresponding flux all group members as predicted by our model 
fits. Since IRAS did not resolve/detect all the members of each group, several of 
the ratios are indicated as upper limits. Filled symbols indicate dynamically ``old'' 
groups.}
\label{fig:result5}
\end{center}
\end{figure}

As it was described in \citet{Bitsakis10}, there are several reasons to expect that the interactions among group members have 
expelled diffuse cold dust along with atomic gas in the intragroup medium. In this preliminary work we had investigated the 
contribution of each galaxy to the total far-IR emission of its group and identified 9 of the 14 groups, in which extended cold 
dust emission may be present. In the present work, using our larger sample of 32 groups and the physical model of \citet{daCunha08} 
we repeat the same analysis. In Fig.~\ref{fig:result5} we plot the fraction of the integrated IRAS flux densities at 60 and 
100$\mu$m for each group as measured by \citet{Allam96} divided by the sum of the flux densities of all group members predicted 
by our SED model fit. The results are presented in Table~\ref{tab:group}. We identify 13 groups (HCG6, 7, 16, 25, 55, 57, 67, 68, 
71, 75, 79, 91 \& 95) where diffuse cold dust emission may be present. These results agree with the previous of \citet{Bitsakis10} 
with HCG37 be the only exception. Moreover, 10 of these groups are dynamically ``old". This is in agreement with the notion that 
dynamically ``old" groups are more evolved  and their members have experienced more encounters enriching the intragroup medium. 
It is possible that part of the apparent far-IR excess is due to the uncertainties introduced by our SED fits. Only far-infrared 
imaging with the Herschel Space Telescope will enable us to reliably characterize the composition and structure of the intragroup 
medium as well as study the evolution of the cold dust in these systems.

\section{Conclusions}
In this paper we have presented the first multi-wavelength analysis, from the UV to the infrared, of the 135 galaxies which are members of 
32 Hickson Compact Groups and studied the influence of the group environment in their evolution. Based on the theoretical modelling of their 
SEDs we conclude the following:

\begin{itemize}
 
\item The classification of the evolutionary state of HCGs according to the fraction of their early-type members appears to be physical and 
is in general agreement with previous classifications. The study of their properties suggest that dynamically ``old'' groups are more compact 
and have higher velocity dispersions. They also display higher stellar masses than the ``young'' ones, while both have similar HI mass 
distributions. However, ``old'' groups have nearly an order of magnitude larger dynamical masses than ``young'' groups.
 
\item The late-type galaxies in dynamically ``old'' groups display lower sSFRs since the multiple past interactions have already converted 
a fraction of their gas into stars increasing their stellar masses. This is also the main reason why these galaxies show redder NUV/optical 
colors than field spirals. However, there are few spiral galaxies in these groups which display even redder colors. They all have very small SFRs, 
similar to early-type systems. We speculate that tidal interactions must have stripped the gas out of their disk suppressing their star 
formation activity.

\item Most early-type galaxies in  dynamically ``old'' groups, seem to migrate from the star-forming to the quiescent galaxy colors, even 
though a fraction of them ($\sim$25\%) display bluer colors and higher star formation activity than normal field ellipticals possibly due 
to gas accretion from other group members as well as merging of dwarf companions.

\item Late-type galaxies in dynamically ``young'' groups have similar star formation properties to field spirals, as well as in early-stage 
interacting pairs.

\item Even though nearly 46\% of the HCG members have an optically identified AGN, we find no evidence of enhanced AGN activity at any 
stage of the group evolution, or the optical/mid-IR colors of the galaxies.
 
\item Our analysis suggests that the reported lower density of galaxies in the IRAC color-color diagram is caused by the morphological 
natural bimodality of galaxies and it is similar to what is also observed in the UV-optical colors.

\item Our SED model suggests that in 13 groups, 10 of which are dynamically ``old", there is diffuse cold dust in the intragroup medium. 

\item There are 10 galaxies with mid-IR colors, SEDs, and star formation properties which are inconsistent with their optical classification 
as early-type systems. We suggest that 7 of them are likely nearly edge-on late-type galaxies which were misclassified due to their small 
size and dust extinction.

\end{itemize}

\begin{acknowledgements}  
TB, VC, and TDS would like to acknowledge partial support from the FP7-REGPOT 206469 grant. We would like to thank H.S. Hwang, 
D. Elbaz and P.-A. Duc (CEA/Saclay, France) for useful discussions which improved the paper, as well as U. Lisenfeld (Univ. of 
Granada) for providing the CO data for a number of groups prior to publication.
\end{acknowledgements} 



\pagebreak
\clearpage
\twocolumn
\begin{table}
\caption{Summary of available observations.}
\begin{center}
\begin{tabular}{lcl}
\hline\hline
Filters & No of galaxies & Comments\\
\hline
GALEX FUV, NUV & 105, 135 & AIS, MIS, and Guest Investigator maps.\\
B, R & 129 & from \citet{Hickson82}.\\
SDSS & 79 & Model magnitudes of all Sloan filters ({\it ugriz}) DR7.\\
J, H, Ks & 128 & 63 galaxies from Palomar Obs., 4 from Skinakas Obs., and 68 from 2MASS. \\
Spitzer/IRAC & 135 & No maps at 3.6 and 5.8$\mu$m for HCG44b. No maps at 4.5 and 8.0$\mu$m for HCG68c.\\
Spitzer/MIPS 24$\mu$m & 124 & No maps or undetected: HCG25f, 40f, 57g, 68d, 68e, 75b, 75e, 92d, 92e, 97b and 97e.\\
IRAS 60, 100$\mu$m & 31 & from \citet{Allam96}.\\
AKARI FIS & 26 & AKARI FIS source catalogue includes the 65, 90, 140, and 160$\mu$m photometry.\\
\hline
Optical Spectra & 94 & from \citet{Martinez10}, \citet{Shimada00}, \citet{Hao05} and \citet{Veron06}.\\
\hline
\hline
\end{tabular}
\label{tab:sample}
\end{center}
\end{table}

\pagebreak
\clearpage

\vspace*{0.1cm}
\begin{sidewaystable}
\centering
\caption{The UV to IR photometry of our sample.}
\begin{tabular}{cccccccccccccc}
\hline\hline
 HCG & Morphology & Nuclear & Distance & FUV & NUV & B & R & u & g & r & i & z & J \\
 galaxy &  & Classification & Mpc & mJy & mJy & mJy & mJy & mJy & mJy & mJy & mJy & mJy & mJy \\
\hline
  37a & E7 & dLINER & 95.8 & 0.10$\pm$0.01 & 0.38$\pm$0.04 & 11.2$\pm$0.56 & 44.5$\pm$2.23 & 3.26$\pm$0.09 & 18.0$\pm$0.54 & 38.0$\pm$1.14 & 54.5$\pm$1.63 & 71.8$\pm$2.15 & 80.7$\pm$2.42 \\
  37b & Sbc & LINER & 95.8 & 0.04$\pm$0.04 & 0.11$\pm$0.01 & 2.58$\pm$0.13 & 13.7$\pm$0.69 & 0.58$\pm$0.02 & 3.40$\pm$0.10 & 8.87$\pm$0.27 & 12.9$\pm$0.39 & 19.6$\pm$0.59 & 33.2$\pm$1.00 \\
  37c & S0a & LLAGN& 104.7 & 0.01$\pm$0.00 & 0.04$\pm$0.00 & 1.03$\pm$0.05 & 4.92$\pm$0.25 & 0.33$\pm$0.01 & 1.64$\pm$0.05 & 3.56$\pm$1.07 & 5.20$\pm$0.16 & 6.92$\pm$0.21 & 8.10$\pm$0.24 \\
  37d & SBdm & -    & 87.0 & 0.10$\pm$0.01 & 0.15$\pm$0.01 & 1.13$\pm$0.06 & 1.32$\pm$0.07 & 0.31$\pm$0.01 & 1.08$\pm$0.03 & 1.67$\pm$0.05 & 2.13$\pm$0.06 & 2.54$\pm$0.08 & 2.74$\pm$0.08 \\
  37e & E0 & TO     & 91.4 & 0.01$\pm$0.00 & 0.03$\pm$0.00 & 0.65$\pm$0.03 & 1.20$\pm$0.06 & 0.19$\pm$0.00 & 0.76$\pm$0.02 & 1.19$\pm$0.04 & 1.43$\pm$0.04 & 1.59$\pm$0.05 & 4.00$\pm$0.12 \\
\hline
\end{tabular}
\label{tab:flux}
\vskip 1cm
\begin{tabular}{cccccccccccccc}
\hline\hline
H & Ks & 3.6$\mu$m & 4.5$\mu$m & 5.8$\mu$m & 8.0$\mu$m & 24$\mu$m & 60$\mu$m & 100$\mu$m & 65$\mu$m & 90$\mu$m & 140$\mu$m & 160$\mu$m & References \\
mJy & mJy & mJy & mJy & mJy & mJy & mJy & mJy & mJy & mJy & mJy & mJy & mJy &  \\
\hline
79.6$\pm$2.39 & 88.1$\pm$2.64 & 61.7$\pm$1.85 & 37.2$\pm$1.12 & 19.3$\pm$0.58 & 13.4$\pm$0.40 & 4.80$\pm$0.24 & 490$\pm$49 & 2000$\pm$200 & - & - & - & - & a2,b1,c1\\
43.3$\pm$1.30 & 43.6$\pm$1.30 & 22.9$\pm$0.68 & 14.4$\pm$0.43 & 14.5$\pm$0.44 & 29.1$\pm$0.87 & 36.7$\pm$1.84 & 561$\pm$56 & - & 675$\pm$34 & 725$\pm$36 & 2156$\pm$107 & 2445$\pm$122 & a2,b1,c1\\
8.90$\pm$0.27 & 7.40$\pm$0.22 & 3.50$\pm$0.11 & 2.30$\pm$0.07 & 1.80$\pm$0.05 & 2.10$\pm$0.06 & 3.50$\pm$0.18 & 190$\pm$19 & - & - & - & - & - & a2,b1,c1\\
3.45$\pm$0.10 & 3.19$\pm$0.10 & 1.84$\pm$0.06 & 1.14$\pm$0.03 & 2.01$\pm$0.06 & 5.19$\pm$0.16 & 4.90$\pm$0.25 & - & - & - & - & - & - & a2,b1,c1\\
5.18$\pm$0.15 & 4.77$\pm$0.14 & 2.41$\pm$0.07 & 1.48$\pm$0.04 & 1.02$\pm$0.03 & 1.07$\pm$0.03 & 1.80$\pm$0.09 & - & - & - & - & - & - & a1,b1,c1\\
\hline\end{tabular}
\tablefoot{
\tablefoottext{a1,2,3,4}{The nuclear classification obtained from \citet{Martinez10}, \citet{Shimada00}, \citet{Hao05} and \citet{Veron06} respectively.}
\tablefoottext{b1,2,3}{The near-IR photometry from Palomar, 2MASS and Skinakas respectively.}
\tablefoottext{c1,2,3,4,5,6}{The mid-IR photometry from \citet{Bitsakis10}, \citet{Johnson07}, and the Spitzer archive (PIDs: 50764, 159,  40385, 198) respectively.  The complete table is available by the authors upon request.}
}
\end{sidewaystable}

\pagebreak
\clearpage
\twocolumn
\begin{table}
\caption{Observational Parameters of Spitzer Archival Data}
\begin{tabular}{ccccccc}
\hline
\hline
  HCG & R.A. (J200))& Dec. (J2000) & z & IRAC$^{a}$ & MIPS$_{24}$$^{a}$  & Observer \\
  ID &  &  &  & sec & sec& PID \\
\hline
  HCG4 & 00h34m15.9s & -21d26m48s & 0.0269 & 10.0 & 18.3 & 40385 \\
  HCG6 & 00h39m10.1s & -08d23m43s & 0.0379 & 150.0 & 375.4 & 50764 \\
  HCG15 & 02h07m39.0s & +02d08m18s & 0.0228 & 300.0 & 375.4 & 50764\\
  HCG25 & 03h20m43.7s & -01d03m07s & 0.0212 & 300.0 & 375.4 & 50764 \\
  HCG44 & 10h18m00.5s & +21d48m44s & 0.0046 & 360.0 & 917.5 & 159 \\
  HCG67 & 13h49m03.5s & -07d12m20s & 0.0245 & 300.0 & 375.4 & 50764\\
  HCG68 & 13h53m40.9s & +40d19m07s & 0.0080 & 75.0 & 375.4 & 50764\\
  HCG75 & 15h21m33.8s & +21d11m00s & 0.0416 & 150.0 & 375.4 & 50764\\
  HCG82 & 16h28m22.1s & +32d49m25s & 0.0362 & 150.0 & 375.4 & 50764\\
  HCG91 & 22h09m10.4s & -27d47m45s & 0.0238 & 300.0 & 375.4 & 50764 \\
  HCG92 & 22h35m57.5s & +33d57m36s & 0.0215 & 432.0 & 627.5 & 198 \\
  HCG97 & 23h47m22.9s & -02d19m34s & 0.0218 & 300.0 & 375.4 & 50764 \\
  HCG100 & 00h01m20.8s & +13d07m57s & 0.0178 & 300.0 & 375.4 & 50764 \\
\hline
\end{tabular}
\tablefoot{
\tablefoottext{a}{On source integration time for each of the four IRAC filters and MIPS 24$\mu$m filter.}
}
\label{tab:obs}
\end{table}

\pagebreak
\clearpage
\onecolumn
\begin{table}
\caption{The morphological type and derived physical parameters based on the SED modeling of the HCG galaxies}
\begin{tabular}{cccccccccc}
\hline\hline
HCG & Morphological & $\chi^{2}$ & $\tau_{V,obs}$ & $\tau_{V,ISM}$ & $\tau_{V,BC+ISM}$ & M$_{star}$ & SFR & sSFR & L$_{\rm IR}$\\
galaxy &  Type &  &  &  &  & $\times$10$^{9}$M$_{\odot}$ & M$_{\odot}$yr$^{-1}$ & $\times$10$^{-11}$yr$^{-1}$ & $\times$10$^{9}$L$_{\odot}$\\
\hline
  2a$^{*}$ & SBd & 2.38 & 0.11 & 0.05 & 0.67 & 3.86 & 2.18 & 57.81 & 14.45\\
  2b & cI & 2.96 & 0.52 & 0.42 & 1.16 & 3.56 & 1.75 & 48.64 & 32.35\\
  2c & SBc & 1.66 & 0.17 & 0.10 & 0.64 & 4.24 & 0.43 & 10.28 & 3.17\\
  4a$^{*}$ & Sc & 5.47 & 0.36 & 0.16 & 1.60 & 69.18 & 7.29 & 10.52 & 169.8\\
  4b & Sc & 5.22 & 0.13 & 0.11 & 0.63 & 7.62 & 0.41 & 5.37 & 4.09\\
  4d & E4 & 5.96 & 0.53 & 0.51 & 0.94 & 12.88 & 0.96 & 7.41 & 1.46\\
  6a$^{*}$ & S0a & 2.65 & 0.13 & 0.14 & 1.98 & 87.1 & 0.01 & 0.01 & 15.84\\
  6b$^{\clubsuit}$ & Sab & 5.69 & 0.24 & 0.23 & 1.45 & 77.62 & 0.05 & 0.07 & 21.87\\
  6c & E5 & 5.49 & 0.10 & 0.10 & 0.70 & 58.88 & 0.01 & 0.01 & 4.38\\
  6d & Sbc & 1.38 & 0.19 & 0.09 & 0.32 & 3.78 & 0.07 & 2.19 & 6.05\\
  7a$^{*}$ & Sb & 2.07 & 0.61 & 0.57 & 2.27 & 74.13 & 1.65 & 2.19 & 2.40\\
  7b & SB0 & 3.58 & 0.17 & 0.17 & 1.68 & 38.90 & 0.14 & 0.36 & 0.77\\
  7c & SBc & 4.37 & 0.34 & 0.33 & 0.38 & 23.44 & 1.61 & 6.92 & 33.88\\
  7d & SBc & 8.61 & 0.03 & 0.02 & 0.65 & 8.85 & 0.32 & 3.16 & 4.38\\
  15a & Sa & 5.59 & 0.09 & 0.10 & 0.64 & 81.28 & 0.10 & 0.11 & 14.45\\
  15b & E0 & 4.60 & 0.24 & 0.24 & 0.82 & 40.74 & 0.05 & 0.11 & 2.40\\
  15c$^{*}$ & E0 & 8.73 & 0.07 & 0.08 & 0.42 & 107.15 & 0.02 & 0.02 & 3.86\\
  15d & E2 & 2.06 & 0.48 & 0.18 & 1.35 & 24.55 & 0.01 & 0.32 & 8.64\\
  15e & Sa & 7.30 & 0.44 & 0.03 & 0.67 & 25.70 & 0.06 & 0.22 & 3.99\\
  15f & Sbc & 7.65 & 0.81 & 0.14 & 0.30 & 2.07 & 0.28 & 14.03 & 2.67\\
  16a & Sba & 5.95 & 0.92 & 0.85 & 2.65 & 50.12 & 4.14 & 7.76 & 0.63\\
  16b$^{*}$ & Sab & 3.41 & 0.12 & 0.11 & 2.19 & 63.10 & 0.04 & 0.07 & 1.78\\
  16c & Im & 2.11 & 1.09 & 0.99 & 2.25 & 11.75 & 8.18 & 69.5 & 72.4\\
  16d & Im & 6.86 & 1.26 & 1.14 & 2.72 & 30.20 & 1.45 & 4.79 & 2.61\\
  22a$^{*}$ & E2 & 2.19 & 0.10 & 0.09 & 0.31 & 70.79 & 0.01 & 0.07 & 100.0\\
  22b & Sa & 2.55 & 0.24 & 0.17 & 0.38 & 5.52 & 0.01 & 0.06 & 89.12\\
  22c & SBc & 4.02 & 0.22 & 0.08 & 0.26 & 3.37 & 0.19 & 7.76 & 1.48\\
  25a & SBc & 5.86 & 0.29 & 0.54 & 1.13 & 13.18 & 9.61 & 71.94 & 3.48\\
  25b$^{*}$ & SBa & 5.62 & 0.03 & 0.03 & 0.31 & 69.18 & 0.16 & 0.23 & 0.85\\
  25d & S0 & 3.58 & 0.14 & 0.14 & 3.40 & 11.48 & 0.05 & 0.58 & 3.06\\
  25f$^{\spadesuit}$ & S0 & 4.24 & 0.04 & 0.04 & 0.15 & 8.17 & 0.05 & 0.60 & 0.17\\
  37a$^{*}$ & E7 & 5.26 & 0.24 & 0.24 & 0.45 & 128.82 & 0.18 & 0.14 & 1.38\\
  37b & Sbc & 4.15 & 0.74 & 0.73 & 3.63 & 74.13 & 0.58 & 0.78 & 26.91\\
  37c & SBd & 3.86 & 0.73 & 0.71 & 3.56 & 17.38 & 0.08 & 0.46 & 1.41\\
  37d & SBdm & 4.70 & 0.24 & 0.25 & 1.77 & 0.89 & 0.15 & 17.66 & 75.85\\
  37e & E0 & 5.26 & 0.21 & 0.20 & 0.70 & 1.65 & 0.03 & 1.95 & 1.20\\
  38a & Sbc & 3.40 & 1.03 & 0.95 & 3.53 & 46.77 & 2.18 & 4.68 & 23.44\\
  38b$^{*}$ & SBd & 7.82 & 1.06 & 1.17 & 3.07 & 50.12 & 4.60 & 8.51 & 0.34\\
  38c & Im & 2.49 & 1.07 & 1.03 & 2.87 & 23.44 & 1.40 & 5.62 & 0.75\\
  40a$^{*}$ & E3 & 0.85 & 0.57 & 0.57 & 2.10 & 165.96 & 0.22 & 0.13 & 38.90\\
  40b & S0 & 2.30 & 0.89 & 0.87 & 1.65 & 83.18 & 0.36 & 0.43 & 12.88\\
  40c & Sbc & 4.24 & 1.46 & 1.13 & 3.64 & 69.18 & 2.39 & 3.47 & 3.21\\
  40d & SBa & 3.02 & 1.07 & 1.02 & 3.79 & 52.48 & 0.59 & 1.12 & 45.70\\
  40e & Scd & 3.65 & 1.50 & 1.40 & 2.33 & 14.13 & 0.66 & 4.27 & 5.03\\
  40f$^{\spadesuit}$ & E1 & 5.13 & 0.02 & 0.02 & 0.58 & 6.64 & 0.01 & 0.01 & 1.00\\
  42a$^{*}$ & E3 & 3.38 & 0.15 & 0.15 & 1.09 & 245.47 & 0.05 & 0.02 & 2.58\\
  42b & SB0 & 7.92 & 0.16 & 0.15 & 1.09 & 25.12 & 0.01 & 0.01 & 16.21\\
  42c & E2 & 2.66 & 0.46 & 0.15 & 1.87 & 20.89 & 0.02 & 0.09 & 20.89\\
  42d & E2 & 2.54 & 0.35 & 0.03 & 0.67 & 3.52 & 0.01 & 0.13 & 77.62\\
  44a$^{*}$ & Sa & 11.89 & 0.34 & 0.34 & 2.29 & 36.31 & 0.07 & 0.20 & 7.62\\
  44b & E2 & 7.09 & 0.38 & 0.38 & 0.66 & 19.95 & 0.03 & 0.17 & 17.78\\
  44c & SBc & 3.93 & 0.29 & 0.29 & 0.54 & 5.46 & 0.11 & 2.09 & 22.90\\
  44d & Sd & 8.52 & 0.31 & 0.31 & 0.57 & 0.64 & 0.63 & 98.17 & 7.53\\
  47a$^{*}$ & Sbc & 10.46 & 0.54 & 0.53 & 0.76 & 95.5 & 3.07 & 3.16 & 1.74\\
  47b & E3 & 9.66 & 0.34 & 0.33 & 0.57 & 75.86 & 0.30 & 0.39 & 0.70\\
  47c & Scd & 3.33 & 1.04 & 1.01 & 3.27 & 15.14 & 0.93 & 6.03 & 43.65\\
  47d & Sdm & 3.63 & 0.14 & 0.13 & 0.22 & 16.98 & 0.48 & 2.75 & 104.71\\
  54a$^{*}$ & Sdm & 7.78 & 1.84 & 1.84 & 3.34 & 0.39 & 3.07 & 16.53 & 34.67\\
  54b & Im & 5.67 & 1.41 & 1.31 & 1.74 & 0.07 & 0.20 & 49.26 & 0.78\\
  54c & Im & 2.35 & 1.13 & 0.61 & 1.19 & 0.02 & 0.08 & 23.44 & 40.73\\
  54d & Im & 4.36 & 0.98 & 0.30 & 2.48 & 0.01 & 0.03 & 32.51 & 33.88\\
  55a & E0 & 5.80 & 0.36 & 0.35 & 0.81 & 93.33 & 0.40 & 0.42 & 64.56\\
  55b & S0 & 4.03 & 0.26 & 0.17 & 1.87 & 70.79 & 0.02 & 0.03 & 36.30\\
  55c$^{*}$ & E3 & 1.68 & 0.94 & 1.01 & 2.96 & 112.2 & 1.17 & 1.05 & 14.45\\
  55d & E2 & 2.89 & 0.18 & 0.18 & 0.98 & 30.20 & 0.08 & 0.26 & 0.07\\
  56a & Sc & 4.68 & 0.52 & 0.51 & 0.54 & 17.78 & 0.70 & 3.89 & 12.58\\
  56b$^{*}$$^{\clubsuit}$ & SB0 & 17.21 & 1.34 & 1.29 & 2.75 & 34.67 & 1.69 & 4.79 & 2.17\\
  56c & S0 & 8.28 & 0.22 & 0.03 & 0.40 & 26.30 & 0.05 & 0.20 & 4.69\\
\hline
\end{tabular}
\tablefoot{
\tablefoottext{*}{Galaxies marked with a star are the most massive of each group and were used in the calculations of the dynamical masses 
in Table 5.}
\tablefoottext{\spadesuit}{ Galaxies with no 24$\mu$m data. The SFRs, sSFRs and L$_{\rm IR}$ of these systems are not well constrained.}
\tablefoottext{\clubsuit}{In these galaxies the presence of an AGN into their nucleus dominates their mid-IR emission.}
}
\label{tab:sfrs}
\end{table}

\begin{table*}
\begin{tabular}{cccccccccc}
\hline\hline
HCG & Morphological & $\chi^{2}$ & $\tau_{V,obs}$ & $\tau_{V,ISM}$ & $\tau_{V,BC+ISM}$ & M$_{star}$ & SFR & sSFR & L$_{\rm IR}$\\
galaxy &  Type &  &  &  &  & $\times$10$^{9}$M$_{\odot}$ & M$_{\odot}$yr$^{-1}$ & $\times$10$^{-11}$yr$^{-1}$ & $\times$10$^{9}$L$_{\odot}$\\
\hline
  56d & S0 & 5.05 & 1.01 & 0.98 & 1.69 & 10.23 & 0.42 & 4.07 & 0.08\\
  56e & S0 & 4.91 & 0.32 & 0.31 & 0.71 & 4.29 & 0.23 & 5.37 & 4.04\\
  57a$^{*}$ & Sbc & 7.75 & 0.28 & 0.29 & 1.20 & 177.83 & 0.76 & 0.43 & 4.33\\
  57b & SBb & 4.15 & 0.28 & 0.28 & 0.52 & 83.18 & 0.49 & 0.60 & 1.43\\
  57c & E3 & 4.00 & 0.06 & 0.05 & 0.66 & 87.10 & 0.01 & 0.01 & 1.80\\
  57d & SBc & 2.31 & 0.22 & 0.18 & 1.64 & 13.80 & 1.77 & 12.22 & 46.77\\
  57e & S0a & 4.74 & 0.89 & 0.87 & 1.86 & 91.20 & 0.19 & 0.20 & 11.48\\
  57f & E4 & 3.28 & 0.40 & 0.33 & 1.96 & 52.48 & 0.03 & 0.05 & 27.54\\
  57g$^{\spadesuit}$ & SB0 & 2.66 & 0.09 & 0.09 & 0.26 & 44.67 & 0.05 & 0.10 & 3.40\\
  57h & SBb & 6.18 & 0.30 & 0.21 & 2.13 & 12.59 & 0.09 & 0.71 & 3.36\\
  59a$^{*}$ & Sa & 19.03 & 0.28 & 0.25 & 2.06 & 14.79 & 0.06 & 0.43 & 4.75\\
  59b & E0 & 7.77 & 0.03 & 0.03 & 0.59 & 7.19 & 0.01 & 0.06 & 0.88\\
  59c & Sc & 1.92 & 0.23 & 0.11 & 0.63 & 1.34 & 0.10 & 8.13 & 0.77\\
  59d & Im & 6.48 & 1.47 & 0.16 & 1.81 & 0.78 & 0.57 & 72.78 & 0.65\\
  61a$^{*}$ & S0a & 4.37 & 0.63 & 0.62 & 3.08 & 89.13 & 0.16 & 0.17 & 11.22\\
  61c & Sbc & 2.30 & 1.30 & 1.23 & 3.74 & 45.71 & 1.79 & 3.80 & 2.99\\
  61d & S0 & 1.58 & 0.02 & 0.02 & 0.29 & 23.44 & 0.04 & 0.17 & 0.33\\
  62a$^{*}$ & E3 & 1.74 & 0.16 & 0.16 & 1.33 & 91.20 & 0.03 & 0.03 & 0.55\\
  62b & S0 & 4.60 & 0.10 & 0.10 & 0.70 & 35.48 & 0.01 & 0.01 & 0.68\\
  62c & S0 & 5.48 & 0.17 & 0.17 & 0.69 & 22.39 & 0.02 & 0.07 & 22.90\\
  62d & E2 & 4.89 & 0.39 & 0.39 & 0.68 & 5.92 & 0.03 & 0.48 & 7.03\\
  67a$^{*}$ & E1 & 6.70 & 0.12 & 0.12 & 0.46 & 245.47 & 0.02 & 0.07 & 2.09\\
  67b & Sc & 8.25 & 1.13 & 1.07 & 2.23 & 23.99 & 1.13 & 4.47 & 1.22\\
  67c & Scd & 4.25 & 0.58 & 0.14 & 0.51 & 14.45 & 0.55 & 3.98 & 0.50\\
  67d & S0 & 0.19 & 0.10 & 0.06 & 0.79 & 14.79 & 0.05 & 0.32 & 23.98\\
  68a & S0 & 14.93 & 0.02 & 0.02 & 0.09 & 67.61 & 0.31 & 0.46 & 4.48\\
  68b & E2 & 6.05 & 0.13 & 0.11 & 2.28 & 60.26 & 0.06 & 0.09 & 42.65\\
  68c$^{*}$ & SBbc & 4.95 & 0.36 & 0.36 & 0.96 & 70.79 & 0.05 & 0.07 & 3.73\\
  68d$^{\spadesuit}$ & E3 & 6.22 & 0.14 & 0.05 & 1.15 & 9.06 & 0.01 & 0.10 & 54.95\\
  68e$^{\spadesuit}$ & S0 & 3.83 & 0.07 & 0.03 & 0.59 & 5.71 & 0.01 & 0.06 & 10.23\\
  71a$^{*}$ & Sbc & 5.22 & 0.25 & 0.16 & 0.28 & 69.18 & 1.65 & 2.34 & 77.62\\
  71b & SB0 & 7.41 & 1.17 & 1.13 & 3.23 & 40.74 & 1.34 & 3.31 & 1.30\\
  71c & Sbc & 4.74 & 0.18 & 0.17 & 0.48 & 7.28 & 0.53 & 7.24 & 15.48\\
  75a$^{*}$ & E4 & 6.84 & 0.14 & 0.14 & 1.39 & 204.17 & 0.26 & 0.13 & 4.18\\
  75b$^{\spadesuit}$ & Sb & 8.62 & 0.03 & 0.03 & 0.04 & 33.11 & 0.01 & 0.01 & 23.44\\
  75c & S0 & 6.18 & 0.21 & 0.18 & 0.46 & 56.23 & 0.02 & 0.06 & 13.80\\
  75d & Sd & 4.39 & 0.45 & 0.29 & 1.09 & 23.44 & 0.65 & 2.75 & 1.80\\
  75e$^{\spadesuit}$ & Sa & 1.72 & 0.01 & 0.04 & 0.24 & 45.71 & 0.06 & 0.14 & 19.05\\
  79a$^{*}$ & Sa & 6.76 & 0.66 & 0.65 & 1.85 & 36.31 & 0.10 & 0.26 & 30.90\\
  79b & S0 & 5.90 & 0.99 & 0.77 & 2.14 & 34.67 & 0.14 & 0.35 & 9.59\\
  79c & S0 & 6.83 & 0.44 & 0.43 & 1.16 & 4.49 & 0.01 & 0.28 & 1.61\\
  79d & Sdm & 3.36 & 1.34 & 1.15 & 2.13 & 0.74 & 3.74 & 48.08 & 1.98\\
  82a$^{*}$ & E3 & 5.11 & 0.39 & 0.38 & 0.70 & 208.93 & 0.07 & 0.03 & 8.35\\
  82b & SBa & 4.75 & 0.78 & 0.76 & 1.72 & 123.03 & 0.15 & 0.12 & 0.12\\
  82c & Im & 4.86 & 1.04 & 0.99 & 3.94 & 33.88 & 3.25 & 9.55 & 0.86\\
  82d & S0a & 8.11 & 0.05 & 0.05 & 1.15 & 28.18 & 0.03 & 0.10 & 54.95\\
  91a$^{*}$ & SBc & 7.59 & 0.41 & 0.39 & 0.56 & 91.2 & 10.16 & 11.02 & 26.91\\
  91b & Sc & 2.41 & 1.01 & 0.88 & 1.80 & 27.54 & 5.22 & 18.49 & 53.70\\
  91c & Sc & 4.73 & 0.16 & 0.08 & 0.65 & 14.79 & 3.22 & 21.48 & 0.41\\
  91d & SB0 & 0.95 & 0.40 & 0.04 & 0.39 & 22.39 & 0.26 & 1.10& 4.33\\
  92a & Sc & 3.74 & 0.23 & 0.23 & 0.65 & 11.75 & 0.79 & 6.76 & 1.40\\
  92b & SBbc & 5.97 & 0.71 & 0.62 & 0.81 & 38.02 & 2.23 & 5.89 & 1.84\\
  92c$^{*}$$^{\clubsuit}$ & SBbc & 5.05 & 2.27 & 2.20 & 3.60 & 39.81 & 26.42 & 64.86 & 1.48\\
  92d$^{\spadesuit}$ & E2 & 5.94 & 0.35 & 0.08 & 0.32 & 34.67 & 0.12 & 0.35 & 7.19\\
  92e$^{\spadesuit}$ & E4 & 3.72 & 0.37 & 0.01 & 0.07 & 64.57 & 0.01 & 0.01 & 43.66\\
  95a$^{*}$ & Sbc & 8.10 & 0.87 & 0.83 & 1.73 & 131.83 & 2.25 & 1.70 & 5.78\\
  95c & Sbc & 5.08 & 1.66 & 1.47 & 3.06 & 16.6 & 14.52 & 87.5 & 0.89\\
  95d & SB0 & 2.25 & 0.64 & 0.61 & 1.41 & 45.71 & 0.96 & 2.14 & 1.22\\
  97a$^{*}$ & E5 & 2.89 & 0.03 & 0.03 & 0.04 & 81.28 & 0.01 & 0.02 & 2.96\\
  97b$^{\spadesuit}$ & Sc & 1.95 & 0.78 & 0.44 & 0.99 & 16.98 & 0.55 & 2.75 & 11.22\\
  97c & Sa & 6.61 & 0.14 & 0.14 & 2.05 & 23.44 & 0.02 & 0.07 & 0.23\\
  97d & E1 & 1.46 & 0.03 & 0.02 & 0.13 & 47.86 & 0.04 & 0.09 & 0.20\\
  97e$^{\spadesuit}$ & S0a & 5.10 & 0.09 & 0.02 & 0.21 & 9.82 & 0.02 & 0.17 & 15.84\\
  100a$^{*}$ & S0a & 2.54 & 0.25 & 0.20 & 1.53 & 61.66 & 1.05 & 1.70 & 60.25\\
  100b & S & 3.56 & 0.52 & 0.50 & 2.08 & 5.65 & 0.34 & 5.89 & 4.86\\
  100c & S & 3.78 & 1.54 & 0.45 & 1.80 & 2.44 & 1.67 & 67.92 & 5.98\\
  100d & Scd & 1.96 & 0.29 & 0.08 & 0.69 & 0.57 & 0.11 & 20.28 & 10.96\\
\hline\end{tabular}
\tablefoot{
\tablefoottext{*}{Galaxies marked with a star are the most massive of each group and were used in the calculations of the dynamical masses 
in Table 5.}
\tablefoottext{\spadesuit}{Galaxies where no 24$\mu$m observations were available and so their SFRs, sSFRs and L$_{IR}$ are not well constrained.}
\tablefoottext{\clubsuit}{In these galaxies the presence of an AGN into their nucleus dominates their IR emission.}
}
\end{table*}

\pagebreak
\clearpage


\begin{table}
\caption{Derived physical parameters based on the SED modelling of the interacting pair galaxy sample} 
\begin{tabular}{ccccccccc}
\hline\hline
Galaxy & z & $\tau_{V,obs}$ & $\tau_{V,ISM}$ & $\tau_{V,BC+ISM}$ & M$_{star}$ & SFR & sSFR & L$_{\rm IR}$\\
 &  &  &  &  & $\times$10$^{9}$M$_{\odot}$ & M$_{\odot}$yr$^{-1}$ & $\times$10$^{-11}$yr$^{-1}$ & $\times$10$^{9}$L$_{\odot}$\\
\hline
  Arp24a & 0.0069 & 0.21 & 0.06 & 0.59 & 0.05 & 0.05 & 74.47 & 0.17\\
  Arp24b & 0.0069 & 0.46 & 0.49 & 1.29 & 0.68 & 5.22 & 948.4 & 48.98\\
  Arp34a & 0.0157 & 0.32 & 0.16 & 0.81 & 4.90 & 1.50 & 30.20 & 12.3\\
  Arp34b & 0.0157 & 0.62 & 0.21 & 1.26 & 30.20 & 0.34 & 1.23 & 6.64\\
  Arp65a & 0.0179 & 0.95 & 0.90 & 2.81 & 34.67 & 1.12 & 3.16 & 39.81\\
  Arp65b & 0.0179 & 0.23 & 0.20 & 1.13 & 19.50 & 0.35 & 1.82 & 5.78\\
  Arp72a & 0.0109 & 0.89 & 0.17 & 0.96 & 2.28 & 3.98 & 165.9 & 30.9\\
  Arp72b & 0.0109 & 0.44 & 0.06 & 0.34 & 0.21 & 0.33 & 164.8 & 2.44\\
  Arp82a & 0.0136 & 0.23 & 0.26 & 1.85 & 18.62 & 2.39 & 12.59 & 29.51\\
  Arp82b & 0.0136 & 0.66 & 0.18 & 0.93 & 3.09 & 0.67 & 23.01 & 6.56\\
  Arp84a & 0.0115 & 1.12 & 1.16 & 2.47 & 9.33 & 3.56 & 36.31 & 51.29\\
  Arp84b & 0.0115 & 0.40 & 0.48 & 1.73 & 61.66 & 3.82 & 6.31 & 57.54\\
  Arp85a & 0.0015 & 0.31 & 0.48 & 2.07 & 9.12 & 0.05 & 0.60 & 3.61\\
  Arp85b & 0.0015 & 0.45 & 0.47 & 1.68 & 8.71 & 2.39 & 25.82 & 27.54\\
  Arp86a & 0.0172 & 0.29 & 0.34 & 1.79 & 114.82 & 4.09 & 3.63 & 63.1\\
  Arp86b & 0.0172 & 1.45 & 0.88 & 3.03 & 2.95 & 4.14 & 120.23 & 41.69\\
  Arp87a & 0.0235 & 0.86 & 0.90 & 2.00 & 9.77 & 3.98 & 38.90 & 54.95\\
  Arp87b & 0.0235 & 0.26 & 0.22 & 1.14 & 16.60 & 2.28 & 13.80 & 22.91\\
  Arp89a & 0.0068 & 0.21 & 0.17 & 1.44 & 0.66 & 0.16 & 23.28 & 1.31\\
  Arp89b & 0.0068 & 1.26 & 0.63 & 2.02 & 41.69 & 0.16 & 0.40 & 18.62\\
  Arp120a & 0.0026 & 0.25 & 0.25 & 1.75 & 11.20 & 0.09 & 0.08 & 1.75\\
  Arp120b & 0.0026 & 0.44 & 0.38 & 1.52 & 17.80 & 0.16 & 0.91 & 5.4\\
  Arp181a & 0.0315 & 0.72 & 0.29 & 1.20 & 97.72 & 2.24 & 2.40 & 37.15\\
  Arp181b & 0.0315 & 0.96 & 0.86 & 2.00 & 52.48 & 3.33 & 6.31 & 81.28\\
  Arp202a & 0.0102 & 1.00 & 0.30 & 1.04 & 0.66 & 1.67 & 234.4 & 11.75\\
  Arp202b & 0.0102 & 0.87 & 0.17 & 0.75 & 0.89 & 1.40 & 363.1 & 29.51\\
  Arp242a & 0.0220 & 1.03 & 0.93 & 3.00 & 43.65 & 3.03 & 7.24 & 50.12\\
  Arp242b & 0.0220 & 0.14 & 0.23 & 1.25 & 39.81 & 1.12 & 2.88 & 12.3\\
  Arp244a & 0.0054 & 0.29 & 0.30 & 1.56 & 26.3 & 4.45 & 16.29 & 54.95\\
  Arp244b & 0.0054 & 0.99 & 0.81 & 2.60 & 89.13 & 2.44 & 3.80 & 83.18\\
  Arp271a & 0.0087 & 0.08 & 0.19 & 0.96 & 12.88 & 7.19 & 54.95 & 58.88\\
  Arp271b & 0.0087 & 0.43 & 0.29 & 1.54 & 6.03 & 3.65 & 61.66 & 32.36\\
  Arp280a & 0.0024 & 0.93 & 0.72 & 2.00 & 0.02 & 0.15 & 916.2 & 1.57\\
  Arp280b & 0.0024 & 0.08 & 0.05 & 2.62 & 2.09 & 0.17 & 8.13 & 1.52\\
  Arp282a & 0.0154 & 0.49 & 0.48 & 2.20 & 85.11 & 0.42 & 0.54 & 19.05\\
  Arp282b & 0.0154 & 0.45 & 0.26 & 1.31 & 7.24 & 0.43 & 6.03 & 5.65\\
  Arp283a & 0.0055 & 0.07 & 0.11 & 0.82 & 1.23 & 0.17 & 15.38 & 1.52\\
  Arp283b & 0.0055 & 1.19 & 1.11 & 2.91 & 12.02 & 1.91 & 15.85 & 35.48\\
  Arp284a & 0.0092 & 0.80 & 0.10 & 0.55 & 0.81 & 0.07 & 8.13 & 0.56\\
  Arp284b & 0.0092 & 1.14 & 0.49 & 2.12 & 2.04 & 6.46 & 275.4 & 57.54\\
  Arp285a & 0.0087 & 0.76 & 0.81 & 2.32 & 18.62 & 1.14 & 5.75 & 26.3\\
  Arp285b & 0.0087 & 0.31 & 0.41 & 1.94 & 8.71 & 0.85 & 9.55 & 11.48\\
  Arp290a & 0.0121 & 0.29 & 0.21 & 1.41 & 29.51 & 0.44 & 1.51 & 5.92\\
  Arp290b & 0.0121 & 1.01 & 0.65 & 2.58 & 28.18 & 0.02 & 0.09 & 7.19\\
  Arp295a & 0.0220 & 1.06 & 0.89 & 3.30 & 7.76 & 11.48 & 144.5 & 114.82\\
  Arp295b & 0.0220 & 0.76 & 0.71 & 2.22 & 26.92 & 0.54 & 2.29 & 30.2\\
  Arp297a & 0.0322 & 0.17 & 0.12 & 0.68 & 31.62 & 1.23 & 3.72 & 12.02\\
  Arp297b & 0.0322 & 0.74 & 0.58 & 1.67 & 5.62 & 0.63 & 11.27 & 10.47\\
  Arp298a & 0.0163 & 0.85 & 1.02 & 2.77 & 14.79 & 2.09 & 13.80 & 33.88\\
  NGC4567a & 0.0163 & 1.92 & 1.14 & 3.35 & 114.82 & 28.84 & 25.12 & 1047.13\\
  NGC4567b & 0.0074 & 1.06 & 0.88 & 3.05 & 58.88 & 6.03 & 11.22 & 138.04\\
\hline
\end{tabular}
\label{tab:intpairs}
\tablefoot{
We calculate the two components of $\tau_{V}$ using: $\tau_{V,ISM}$=$\mu \tau_{V}$ and $\tau_{V,BC}$=(1-$\mu$)$\tau_{V}$.
}
\end{table}

\pagebreak
\clearpage

\begin{table}
\caption{Distance to, virial radius and morphology of the nearest neighbor as well as ``strength'' of interaction for 
the HCG galaxies.} 
\begin{tabular}{ccccc}
\hline\hline
HCG & R $^{a}$& r$_{vir}$ $^{b}$ & $<$R/r$_{vir,nei}$$>$/n $^{c}$ & type of $^{d}$ \\
galaxy & kpc & kpc &   & nearest neighbor \\
\hline
  2a & 24 & 404 & 0.092 & late\\
  2b & 24 & 357 & 0.108 & late\\
  2c & 76 & 255 & 0.673 & late\\
  4a & 62 & 662 & 0.085 & late\\
  4b & 28 & 467 & 0.041 & early\\
  4d & 28 & 399 & 0.070 & late\\
  6a & 37 & 745 & 0.047 & early\\
  6b & 13 & 571 & 0.023 & late\\
  6c & 13 & 528 & 0.025 & late\\
  6d & 35 & 293 & 0.028 & late\\
  7a & 36 & 523 & 0.043 & late\\
  7b & 21 & 458 & 0.038 & late\\
  7c & 76 & 452 & 0.058 & late\\
  7d & 21 & 551 & 0.032 & early\\
  15a & 101 & 736 & 0.055 & early\\
  15b & 64 & 529 & 0.028 & late\\
  15c & 49 & 868 & 0.030 & early\\
  15d & 14 & 507 & 0.034 & late\\
  15e & 64 & 591 & 0.043 & early\\
  15f & 14 & 406 & 0.020 & early\\
  16a & 16 & 563 & 0.032 & late\\
  16b & 16 & 488 & 0.038 & late\\
  16c & 37 & 462 & 0.034 & late\\
  16d & 37 & 611 & 0.047 & late\\
  22a & 37 & 555 & 0.070 & late\\
  22b & 17 & 267 & 0.034 & late\\
  22c & 17 & 285 & 0.034 & late\\
  25a & 57 & 552 & 0.059 & early\\
  25b & 9 & 667 & 0.032 & early\\
  25d & 46 & 334 & 0.041 & late\\
  25f & 9 & 310 & 0.029 & late\\
  37a & 17 & 932 & 0.037 & early\\
  37b & 20 & 455 & 0.039 & late\\
  37c & 29 & 449 & 0.032 & early\\
  37d & 20 & 245 & 0.037 & late\\
  37e & 30 & 150 & 0.037 & late\\
  38a & 87 & 437 & 0.104 & late\\
  38b & 9 & 511 & 0.056 & late\\
  38c & 9 & 399 & 0.060 & late\\
  40a & 14 & 795 & 0.008 & late\\
  40b & 12 & 685 & 0.008 & late\\
  40c & 12 & 774 & 0.007 & early\\
  40d & 20 & 760 & 0.012 & early\\
  40e & 13 & 491 & 0.006 & early\\
  42a & 21 & 998 & 0.034 & early\\
  42b & 81 & 583 & 0.067 & early\\
  42c & 21 & 510 & 0.041 & early\\
  42d & 31 & 288 & 0.031 & early\\
  44a & 26 & 335 & 0.044 & late\\
  44b & 26 & 393 & 0.069 & late\\
  44c & 54 & 348 & 0.078 & late\\
  44d & 26 & 213 & 0.044 & late\\
  47a & 34 & 778 & 0.052 & early\\
  47b & 34 & 554 & 0.034 & late\\
  47c & 15 & 389 & 0.028 & late\\
  47d & 15 & 405 & 0.025 & late\\
  54a & 1 & 147 & 0.007 & late\\
  54b & 1 & 152 & 0.011 & late\\
  54c & 1 & 110 & 0.005 & late\\
  54d & 1 & 87 & 0.006 & late\\
  55a & 23 & 929 & 0.015 & early\\
  55b & 17 & 710 & 0.019 & early\\
  55c & 17 & 790 & 0.015 & early\\
  55d & 23 & 588 & 0.019 & early\\
  56a & 59 & 476 & 0.046 & early\\
\hline
\end{tabular}
\tablefoot{
\tablefoottext{a}{Projected distance to the nearest neighbor.} 
\tablefoottext{b}{Virial radius.}
\tablefoottext{c}{``Strength'' of interaction as described in Section 4.6.}
\tablefoottext{d}{Morphology of the nearest neighbor.} 
}
\label{tab:dynnei}
\end{table}

\begin{table*} 
\begin{tabular}{ccccc}
\hline\hline
HCG & R $^{a}$& r$_{vir}$ $^{b}$ & $<$R/r$_{vir,nei}$$>$/n $^{c}$ & morphology of $^{d}$\\
galaxy & kpc & kpc &   & nearest neighbor \\
\hline
  56b & 31 & 622 & 0.027 & early\\
  56c & 12 & 580 & 0.021 & early\\
  56d & 12 & 402 & 0.022 & early\\
  56e & 27 & 436 & 0.030 & early\\
  57a & 13 & 669 & 0.031 & late\\
  57b & 26 & 748 & 0.020 & early\\
  57c & 26 & 583 & 0.031 & late\\
  57d & 13 & 252 & 0.020 & late\\
  57e & 32 & 516 & 0.028 & late\\
  57f & 86 & 641 & 0.060 & early\\
  57g & 26 & 463 & 0.022 & late\\
  57h & 32 & 421 & 0.025 & early\\
  59a & 13 & 368 & 0.036 & late\\
  59b & 33 & 298 & 0.066 & late\\
  59c & 26 & 232 & 0.046 & late\\
  59d & 13 & 182 & 0.035 & late\\
  61a & 43 & 705 & 0.043 & early\\
  61c & 24 & 805 & 0.033 & early\\
  61d & 24 & 400 & 0.023 & late\\
  62a & 6 & 745 & 0.023 & early\\
  62b & 6 & 500 & 0.025 & early\\
  62c & 19 & 518 & 0.028 & early\\
  62d & 56 & 354 & 0.035 & early\\
  67a & 17 & 1032 & 0.028 & early\\
  67b & 99 & 635 & 0.054 & late\\
  67c & 25 & 624 & 0.031 & early\\
  67d & 17 & 486 & 0.028 & early\\
  68a & 10 & 710 & 0.032 & early\\
  68b & 10 & 629 & 0.030 & early\\
  68c & 33 & 532 & 0.038 & early\\
  68d & 49 & 370 & 0.027 & early\\
  68e & 47 & 337 & 0.037 & early\\
  71a & 68 & 799 & 0.082 & early\\
  71b & 68 & 502 & 0.066 & late\\
  71c & 76 & 431 & 0.064 & early\\
  75a & 3 & 920 & 0.033 & early\\
  75b & 3 & 534 & 0.033 & late\\
  75c & 23 & 498 & 0.032 & late\\
  75d & 23 & 533 & 0.025 & early\\
  75e & 41 & 453 & 0.023 & late\\
  79a & 4 & 168 & 0.006 & late\\
  79b & 8 & 523 & 0.013 & early\\
  79c & 4 & 366 & 0.009 & early\\
  79d & 5 & 229 & 0.009 & late\\
  82a & 47 & 789 & 0.054 & late\\
  82b & 66 & 819 & 0.097 & early\\
  82c & 44 & 171 & 0.028 & early\\
  82d & 44 & 440 & 0.053 & late\\
  91a & 15 & 1047 & 0.039 & early\\
  91b & 82 & 662 & 0.057 & late\\
  91c & 15 & 553 & 0.031 & early\\
  91d & 15 & 527 & 0.032 & late\\
  92a & 5 & 629 & 0.019 & late\\
  92b & 1 & 649 & 0.013 & early\\
  92c & 4 & 925 & 0.020 & late\\
  92d & 1 & 663 & 0.013 & late\\
  92e & 5 & 801 & 0.024 & early\\
  95a & 20 & 914 & 0.032 & late\\
  95c & 20 & 726 & 0.030 & late\\
  95d & 48 & 510 & 0.031 & late\\
  97a & 37 & 757 & 0.031 & early\\
  97b & 112 & 568 & 0.054 & early\\
  97c & 75 & 528 & 0.043 & early\\
  97d & 37 & 682 & 0.034 & early\\
  97e & 41 & 442 & 0.038 & early\\
  100a & 28 & 621 & 0.036 & late\\
  100b & 34 & 506 & 0.056 & early\\
  100c & 43 & 394 & 0.047 & late\\
  100d & 28 & 254 & 0.031 & early\\
\hline
\end{tabular}
\tablefoot{
\tablefoottext{a}{Projected distance to the nearest neighbor.} 
\tablefoottext{b}{Virial radius.}
\tablefoottext{c}{``Strength'' of interaction as described in the text.}
\tablefoottext{d}{Morphology of the nearest neighbor.} 
}
\end{table*}

\pagebreak
\clearpage

\begin{table}
\caption{Observed and predicted 60$\mu$m and 100$\mu$m IRAS fluxes for each 
group, as well as total HI, H$_2$, stellar, and dynamical masses }
\begin{tabular}{ccccccccccccc}
\hline
\hline
HCG & Evolve & HI & Group & f$_{60,pred}$ & f$_{60,obs}$ & f$_{100,pred}$ & f$_{100,obs}$ & log(M$_{\rm HI}$)$^{d}$ & 
 log(M$_{star}$) & log(M$_{\rm H_2}$)$^{e}$ & log(M$_{dyn}$) & (M$_{bar}$/M$_{dyn}$)$^{f}$\\
ID & Class$^{a}$ & Class$^{c}$ & members & Jy & Jy & Jy & Jy & M$_{\odot}$ & M$_{\odot}$ & M$_{\odot}$ & M$_{\odot}$ &\\
\hline
  2 & young & - & 3 & 6.33 & 6.16 & 4.03 & 2.89 & 10.33 & 10.07 & - & 11.67 & 0.07\\
  4 & young$^{b}$ & - & 3 & 8.21 & 8.56 & 4.35 & 4.09 & 10.31 & 10.95 & - & 11.96 & 0.12\\
  6 & old & - & 4 & 0.34 & 0.46 & 0.11 & 0.13 & 9.69 & 11.36 & - & 12.66 & 0.05\\
  7 & young & 1 & 4 & 8.32 & 9.20 & 3.42 & 4.43 & 9.68 & 11.16 & $<$9.88 & 11.64 & 0.35\\
  15 & old & 3 & 6 & 1.29 & 0.38 & 0.94 & 0.13 & 9.41 & 11.14 & $<$9.73 & 13.24 & 0.01\\
  16 & young & 2 & 4 & 35.72 & 62.37 & 24.04 & 25.21 & $>$10.42 & 11.19 & 10.41 & 12.57 & $>$0.06\\
  22 & old & - & 3 & 2.72 & 1.48 & 1.0 & 1.05 & 9.13 & 10.90 & - & 11.33 & 0.38\\
  25 & old & 2 & 4 & 1.4 & 5.21 & 0.68 & 1.74 & 9.90 & 11.00 & $<$9.61 & 11.13 & 0.83\\
  37 & old & 3 & 5 & 2.73 & 2.60 & 1.22 & 0.89 & 9.19 & 11.34 & 9.88 & 13.01 & 0.02\\
  38 & young & - & 3 & 7.09 & 3.60 & 3.16 & 1.63 & 9.69 & 11.08& - & 11.31 & 0.61\\
  40 & old & 3 & 6 & 4.82 & 3.85 & 3.42 & 1.35 & 9.14 & 11.59& 10.07 & 12.29 & 0.21\\
  42 & old & - & 4 & 5.14 & - & 1.66 & 0.0 & 9.40 & 11.47 & - & 13.17 & 0.02\\
  44 & young & 3 & 4 & 15.68 & 14.32 & 8.28 & 5.69 & 9.23 & 10.79 & $<$9.17 & 12.70 & 0.01\\
  47 & young & - & 4 & 1.66 & 1.51 & 0.82 & 1.27 & 9.93 & 11.30 & - & 11.63 & 0.48\\
  54 & young & - & 4 & 2.02 & 1.71 & 0.35 & 0.38 & 8.75 & 8.69 & - & 11.25 & 0.01\\
  55 & old & - & 4 & 1.1 & 1.63 & 0.26 & 0.2 & - & 11.47 & - & 12.53 & 0.09\\
  56 & old & - & 5 & 2.59 & 2.27 & 1.13 & 0.89 & 9.36 & 10.97 & - & 12.81 & 0.02\\
  57 & old & - & 8 & 3.14 & 3.54 & 1.16 & 1.44 & 9.71 & 11.75 & - & 12.58 & 0.15\\
  59 & young & - &4 & 13.3 & 4.83 & 6.91 & 3.98 & 9.49 & 10.38& - & 12.25 & 0.02\\
  61 & old & - &3 & 16.3 & 12.45 & 7.18 & 6.39 & 9.96 & 11.20& - & 12.33 & 0.08\\
  62 & old & - & 4 & 1.45 & 0.72 & 0.29 & 0.13 & $<$9.06 & 11.19& - & 12.85 & $<$0.02\\
  67 & old & 1 & 4 & 2.98 & 2.91 & 1.15 & 2.26 & 10.03 & 11.48& $<$10.04 & 12.16 & 0.21\\
  68 & old & 3 & 5 & 4.97 & 10.45 & 2.08 & 2.74 & 9.62 & 11.33& $<$9.23 & 12.82 & 0.03\\
  71 & young$^{b}$ & - & 3 & 2.75 & 3.69 & 1.29 & 1.84 & - & 11.07& - & 11.55 & 0.33\\
  75 & old & - & 5 & 1.4 & 2.15 & 0.16 & 0.57 & - & 11.56& - & 12.74 & 0.07\\
  79 & old & 1 & 4 & 2.89 & 3.80 & 2.28 & 1.41 & 9.30 & 10.88& $<$9.31 & 11.72 & 0.15\\
  82 & old & - & 4 & 3.92 & 3.31 & 0.96 & 3.47 & $<$9.69 & 11.60& - & 13.04 & $<$0.04\\
  91 & young & 2 & 4 & 8.78 & 6.11 & 5.0 & 2.31 & 10.32 & 11.19& - & 12.67 & 0.04\\
  92 & old & 3 & 5 & 7.15 & 6.57 & 1.34 & 1.39 & 10.02 & 11.28& - & 12.82 & 0.03\\
  95 & old & - & 3 & 9.44 & 2.39 & 2.48 & 6.94 & $>$10.10 & 11.29& - & 12.70 & $>$0.04\\
  97 & old & 3 & 5 & 5.28 & - & 2.12 & 0.0 & 9.10 & 11.25& $<$9.80 & 13.02 & 0.02\\
  100 & young & 2 & 4 & 5.78 & 6.46 & 2.71 & 2.36 & 9.74 & 10.85 & $<$9.37 & 11.16 & 0.54\\
\hline\end{tabular}
\tablefoot{
\tablefoottext{a}{The dynamically ``young'' and ``old'' classification as discussed in Section 4.1 and in \citet{Bitsakis10}.\\} 
\tablefoottext{b}{The original classifications of HCG4 and HCG71 were dynamically ``old'', but changed as explained in Section 4.3.\\} 
\tablefoottext{c}{The classification to Phase 1, 2, and 3, follows the definition of \citet{Verdes01}.\\} 
\tablefoottext{d}{The upper and lower limits, in HI masses, have been taken into account by means of survival analysis.\\}
\tablefoottext{e}{The H$_{2}$ masses were obtained from \citet{Martinez-Badenes11}.\\}
\tablefoottext{f}{Ratio of the total baryonic mass M$_{bar}$=M$_{star}$+M$_{\rm HI}$+M$_{\rm H_2}$ over the dynamical mass M$_{dyn}$.}
}
\label{tab:group}
\end{table}

\newpage
\onecolumn
\appendix
\section{Peculiar early type systems}

As we mentioned in Section 4.4, there are 10 peculiar early-type galaxies which display ``red'' mid-IR colors. It was suggested that they 
could either be misclassified late-type systems or that they are indeed early-type which have increased their gas and dust due  to 
interactions and merging with companion galaxies.  As we discussed in Section 4.2, some of them display high Av, SFR, and sSFR. 
In this section we present all available evidence and we discuss in detail their properties. 

In order to better probe their morphology, we first examine how the old stellar populations, traced by their near-IR emission, as well 
as the warm dust seen in the mid-IR are distributed in each galaxy. In Fig.~\ref{fig:appendix} we present the ``true color'' images 
constructed using the Spitzer/IRAC 3.6, 4.5 and 8.0$\mu$m maps as the blue, green, and red channel respectively. We also include 
contour maps of the Ks emission for these galaxies. Bona fide early-type systems should be relaxed dynamically with an R$^{1/4}$ radial 
profile and elliptical structure in their stellar light emission. Furthermore, they should not display star formation or color gradients 
hence no variation would be expected in their mid-IR ``true color'' images. Indeed, when one observes the ``true color'' image of NGC1404 
(a typical field elliptical galaxy) no obvious mid-IR color gradient is seen while the contours of the Ks-band emission, dominated  by the 
old stellar population are concentric ellipses. HCG40f and HCG68a, also display similar mid-IR colors and profiles with the second having Ks 
contours consistent with a lenticular galaxy. However the rest of the early-type HGC galaxies show either color gradients, or patchy emission 
in their ``true color'' mid-IR images, which indicate possible on-going star formation warming up the dust and exciting the PAH molecules in 
these wavelengths. In particular the galaxies HCG4d, HCG55c, HCG56d, HCG71b and HCG79b display reddish colors which imply stronger 8$\mu$m 
emission. In HCG100a the 8.0$\mu$m emission seems to emerge from two spiral-like structures. Yet when we closely examine the distribution of 
their old stellar population it appears relaxed. With the exception of HCG56b and HCG71b which display a barred-like structure, no other 
system shows obvious signs of spiral or irregular structure, even though it would have been challenging to identify such features in edge-on 
late type systems. 
 
An additional tool which can help us classify these galaxies is their global SEDs and current star formation activity. In 
Fig.~\ref{fig:appendixsed} we present their SEDs and contrast it with the NGC1404, a typical E1  galaxy the SED of which was also fitted 
with the \citet{daCunha08} model. We note that HCG40f and HCG68a display SEDs which are consistent with that of an elliptical galaxy. It 
is very possible that HCG68a has ``red'' mid-IR colors because of the existence of an AGN into its nucleus (see Table~\ref{tab:tengal}. 1). 
HCG40f was not classified by  \citet{Hickson82} but by us \citep{Bitsakis10}, since its morphology and SED were consistent an early-type 
galaxy. However, this galaxy is very close to its neighbor (HCG40d) and it is possible that this could slightly affect its mid-IR colors. 
The SED of HCG56b is strongly affected by the emission of its Sy2 nucleus which dominates its mid-IR colors. Since our model cannot account 
for emission due to an AGN we can not draw a firm conclusion on its nature. Among the galaxies for which nuclear optical spectra were 
available four systems:  HCG56d, HCG68a, HCG79b and HCG100a display signatures of an AGN or a transition object, while HCG56e and HCG71b 
are classified as pure star forming. The remaining of the galaxies display SEDs which are similar with those of late-type galaxies. They 
have strong mid-IR fluxes consistent with the presence of PAH emission. 

With the exception of HCG40f, all nine galaxies have high sSFRs similar to those of late-type systems (see Table~\ref{tab:tengal}. 1). 
More specifically HCG4d, HCG55c, HCG56b, HCG56d, HCG56e, HCG71b and HCG100a have sSFRs which are two orders of magnitude 
larger than what is seen in normal ellipticals, while HCG68a and HCG79b have sSFRs that are just a factor of ten higher. Finally, as 
we mentioned in Section 4.5, if these galaxies were passively evolving they would display ``red'' NUV-r colors ($>$5mag). However, 
the observed NUV-r colors of HCG4d and HCG56b are blue. After correcting for the dust extinction seven of the remaining of the galaxies 
also move to the ``blue cloud''. It is thus possible that these galaxies are dust obscured late-types with some on-going star formation 
activity. As a result they display higher sSFRs,  ``bluer'' colors in the UV-optical and ``redder'' in the mid-IR wavelengths. Even after 
extinction correction HCG40f  remains in the ``red sequence''  while HCG68a is found in the ``green valley''. These galaxies are early-type 
galaxies which display red mid-IR colors probably because they increased their dust content due to interactions with their neighbors 
(HCG40d and HCG68b, respectively).

Taking in account all the evidence mentioned above we suggest that HCG4d, HCG55c, HCG56d, HCG56e, HCG71b, HCG79b and HCG100a 
could be misclassified edge-on late-type systems.\\

\begin{table}
\begin{center}
\caption{Properties of the 10 peculiar early-type HCG galaxies and the typical elliptical NGC\,1404.}
\begin{tabular}{lcccccccccc}
\hline \hline
Galaxy & Type $^{a}$ & Optical & Nuclear & SED & $\tau_{V,obs}$ & SFR & sSFR & (NUV-r)$_{corr}$  & Proposed \\
 &  & feature & type &  type &  & M$_{\odot}$yr$^{-1}$ & $\times$10$^{-11}$yr$^{-1}$ & [mag] & type \\
\hline
HCG\,4d & E4 & - & - & late-type & 0.53 & 0.96 & 7.41 & 2.01 & late-type \\
HCG\,40f$^{b}$ & - & - & - & early-type & 0.02 & 0.01 & 0.01 & 5.11 & early-type \\
HCG\,55c & E3 & - & - & late-type & 0.94 & 1.17 & 1.05 & 2.90 & late-type \\
HCG\,56b$^{c}$ & SB0 & bar & Sy2 & AGN dom. & 1.34 & 1.69 & 4.79 & 2.04 & - \\
HCG\,56d & S0 & - & Sy2 & late-type & 1.01 & 0.42 & 4.07 & 2.81 & late-type \\
HCG\,56e & S0 & - & HII & late-type & 0.89 & 0.19 & 5.37 & 2.78 & late-type \\
HCG\,68a & S0 & - & AGN & early-type & 0.02 & 0.31 & 0.46 & 3.81 & early-type \\
HCG\,71b & SB0 & bar & HII & late-type & 1.17 & 1.34 & 3.31 & 2.74 & late-type \\
HCG\,79b & S0 & - & TO & late-type & 0.99 & 0.10 & 0.35 & 2.95 & late-type \\
HCG\,100a & S0 & - & TO & late-type & 0.25 & 1.05 & 1.70 & 2.24 & late-type \\
\hline
NGC\,1404 & E1 & - & - & early-type & 0.02 & 0.15 & 0.08 & 5.49 & - \\
\hline
\end{tabular}
\tablefoot{
\tablefoottext{a}{Based on \citet{Hickson82}.} 
\tablefoottext{b}{Galaxy HCG40f is not detected at 24$\mu$m, thus its SFR and sSFR are not well constrained.}
\tablefoottext{c}{Galaxy HCG56b emission is dominated by the presence of an AGN.}
}
\end{center}
\label{tab:tengal}
\end{table}


\begin{figure*}
\begin{center}
\includegraphics[scale=1.]{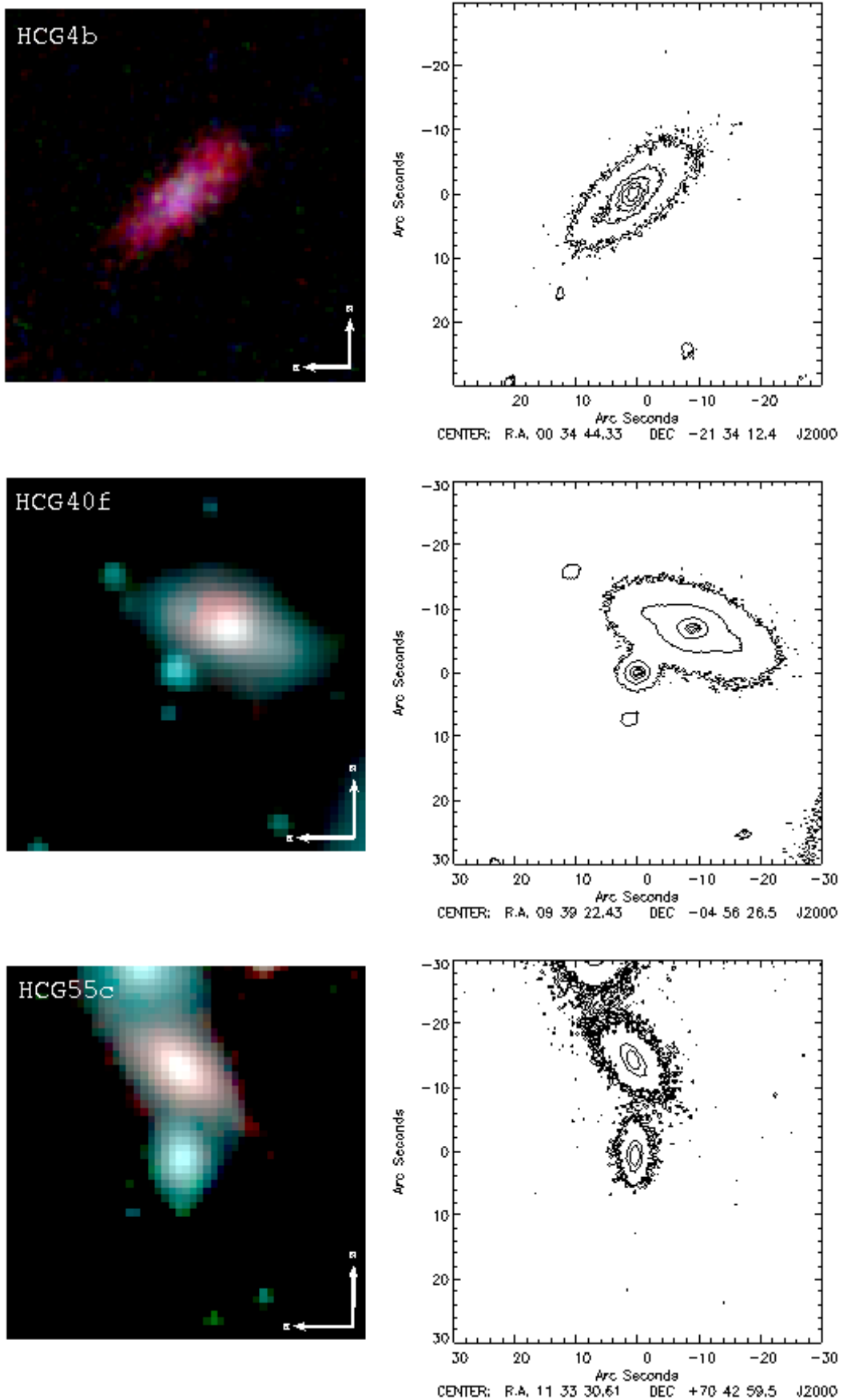}
\caption{{\bf Left column:} The ``true color'' images, created using the 3.6, 4.5 and 8.0$\mu$m Spitzer/IRAC mosaics, 
of the 10 galaxies discussed in Section 4.3. The images are aligned north-east and they are 1.5$\times$1.5 arcmin 
in size. {\bf Right column:} Contour maps of the same fields observed in Ks-band with WIRC at the Palomar 5m telescope. 
The five contours correspond to  3, 5, 8, 10, and 30$\sigma$. Since all images had similar exposure time the noise was 
$\sigma$$\sim$19.6 mag arcsec$^{-2}$.  For two galaxies, HCG68a and HCG100a, not imaged with WIRC we used the shallower 
Ks-band data from 2MASS  ($\sigma$=15.7 mag arcsec$^{-2}$).}
\label{fig:appendix}
\end{center}
\end{figure*}

\begin{figure*}
\setcounter{figure}{0} 
\begin{center}
\includegraphics[scale=1.]{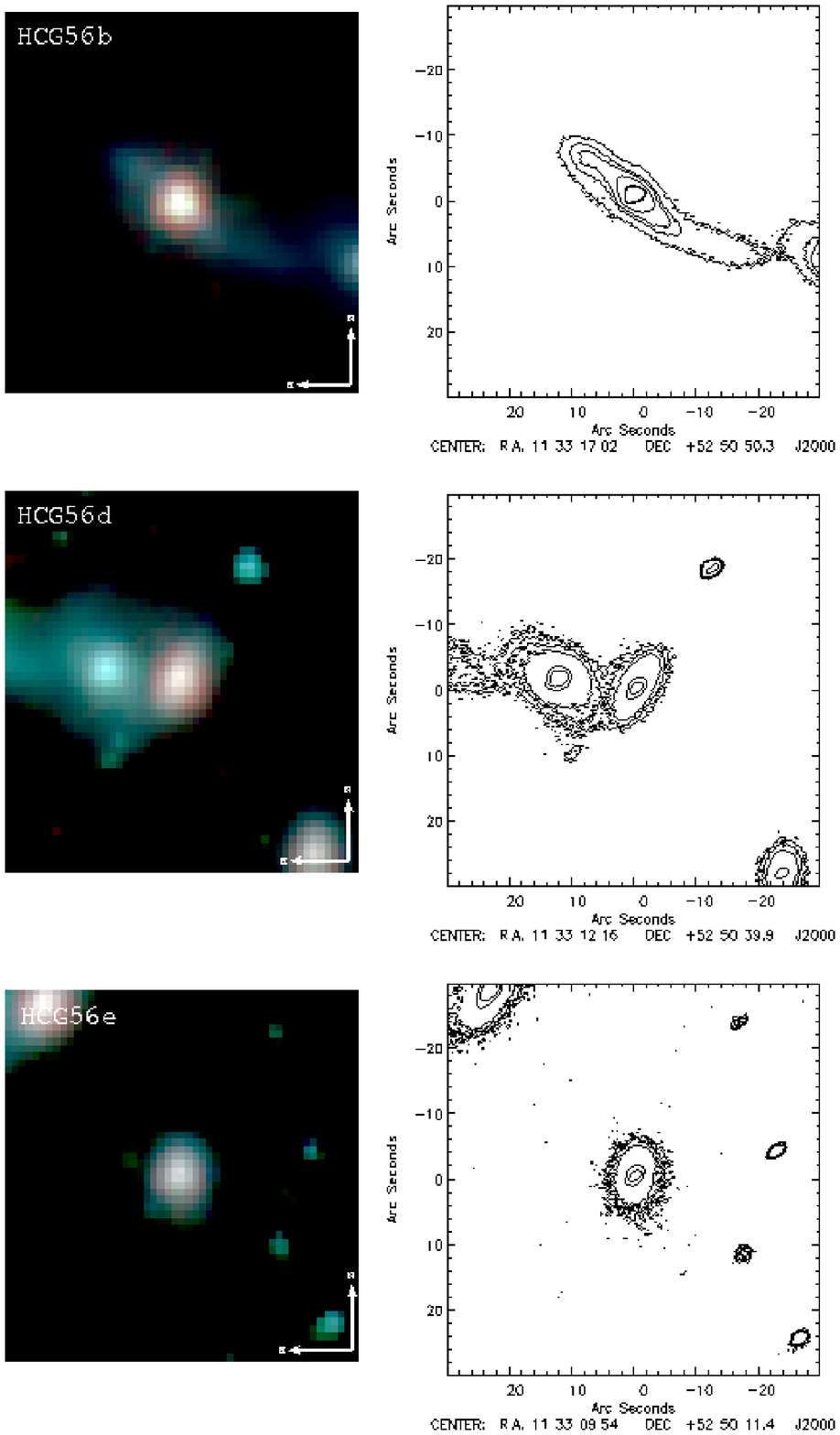}
\caption{Continued}
\end{center}
\end{figure*}

\begin{figure*}
\setcounter{figure}{0} 
\begin{center}
\includegraphics[scale=1.]{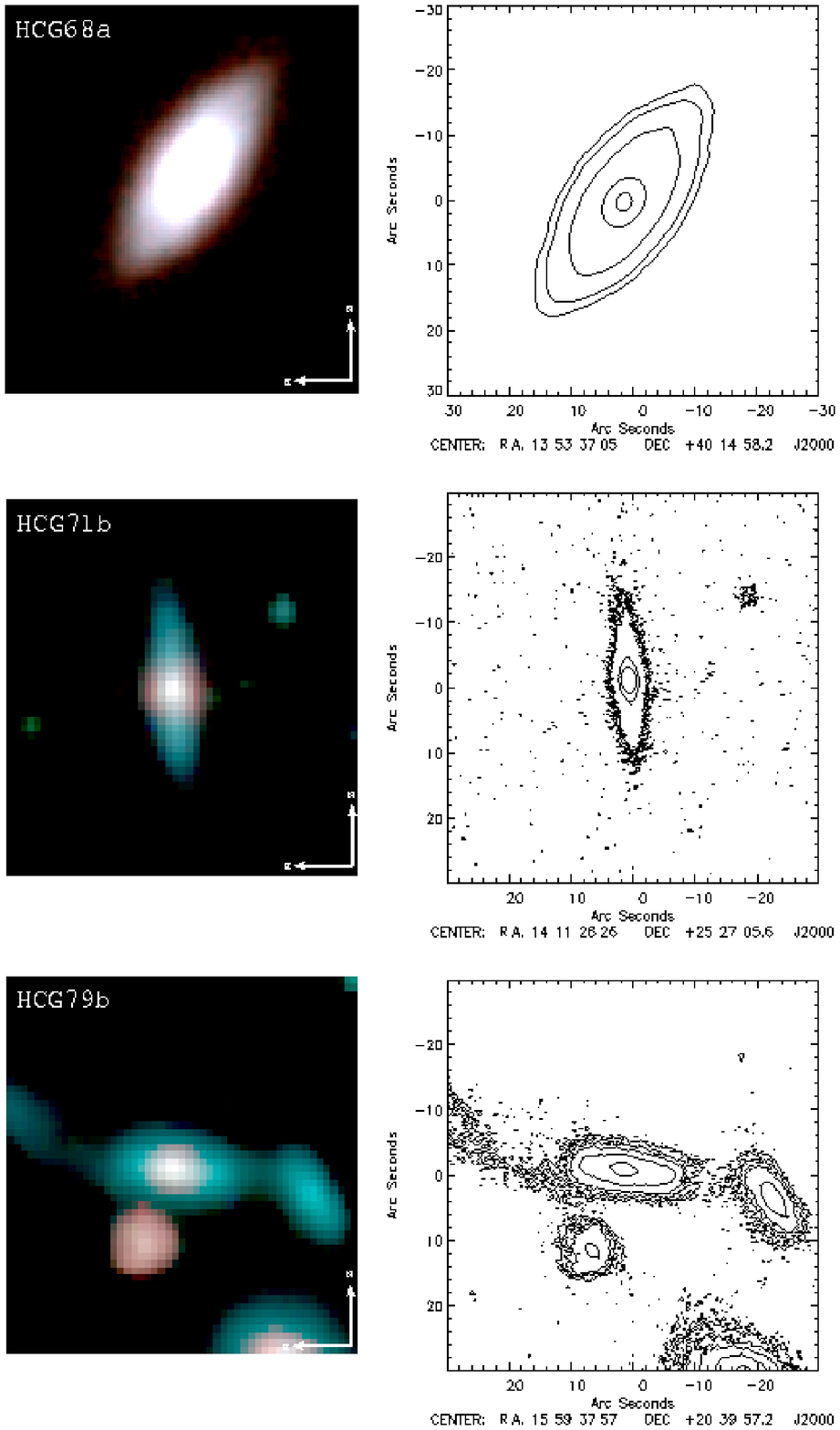}
\caption{Continued}
\end{center}
\end{figure*}

\begin{figure*}
\setcounter{figure}{0}
\begin{center}
\includegraphics[scale=1.]{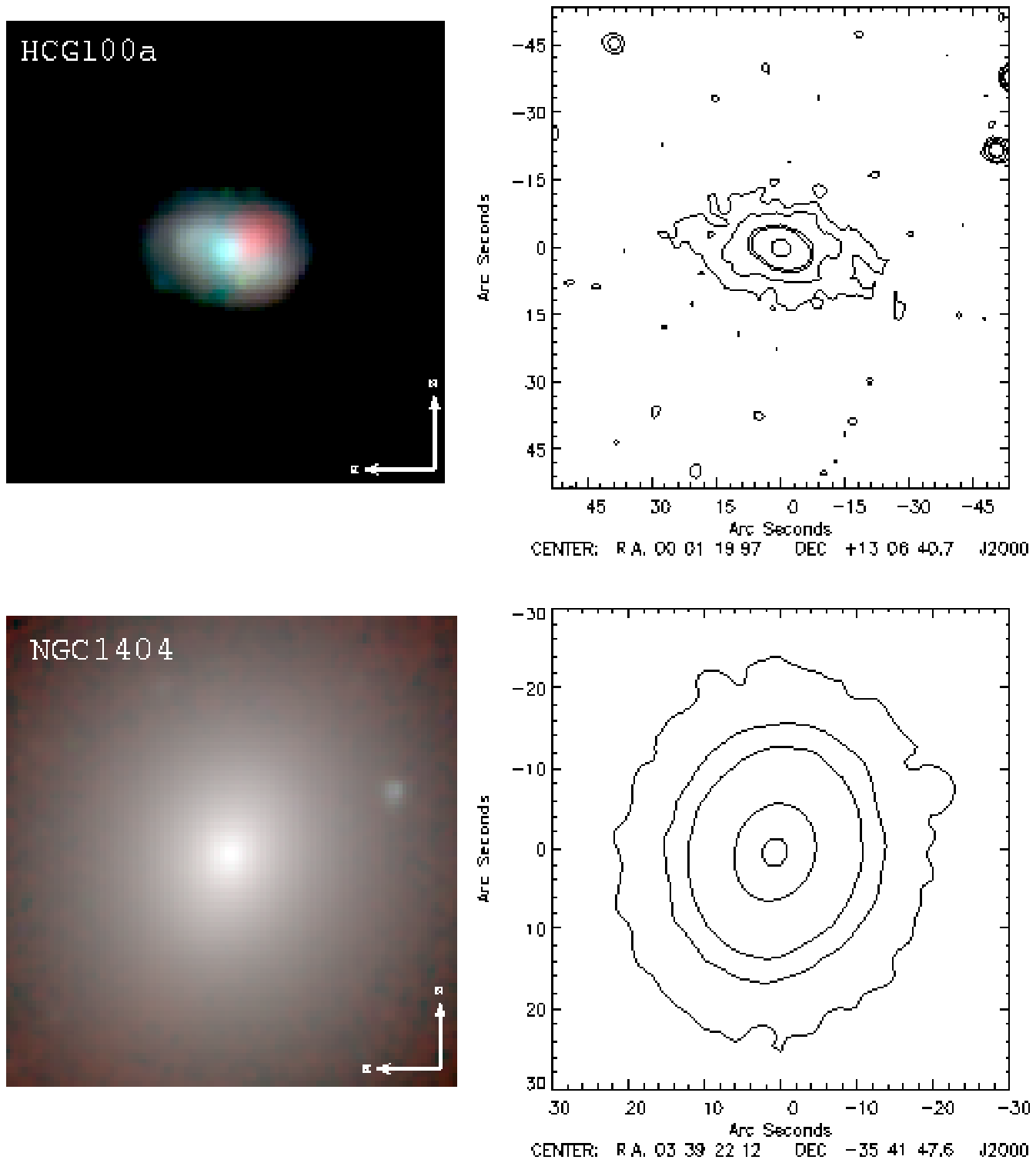}
\caption{Continued}
\end{center}
\end{figure*}


\onecolumn
\begin{figure}
\begin{center}
\includegraphics[scale=0.85]{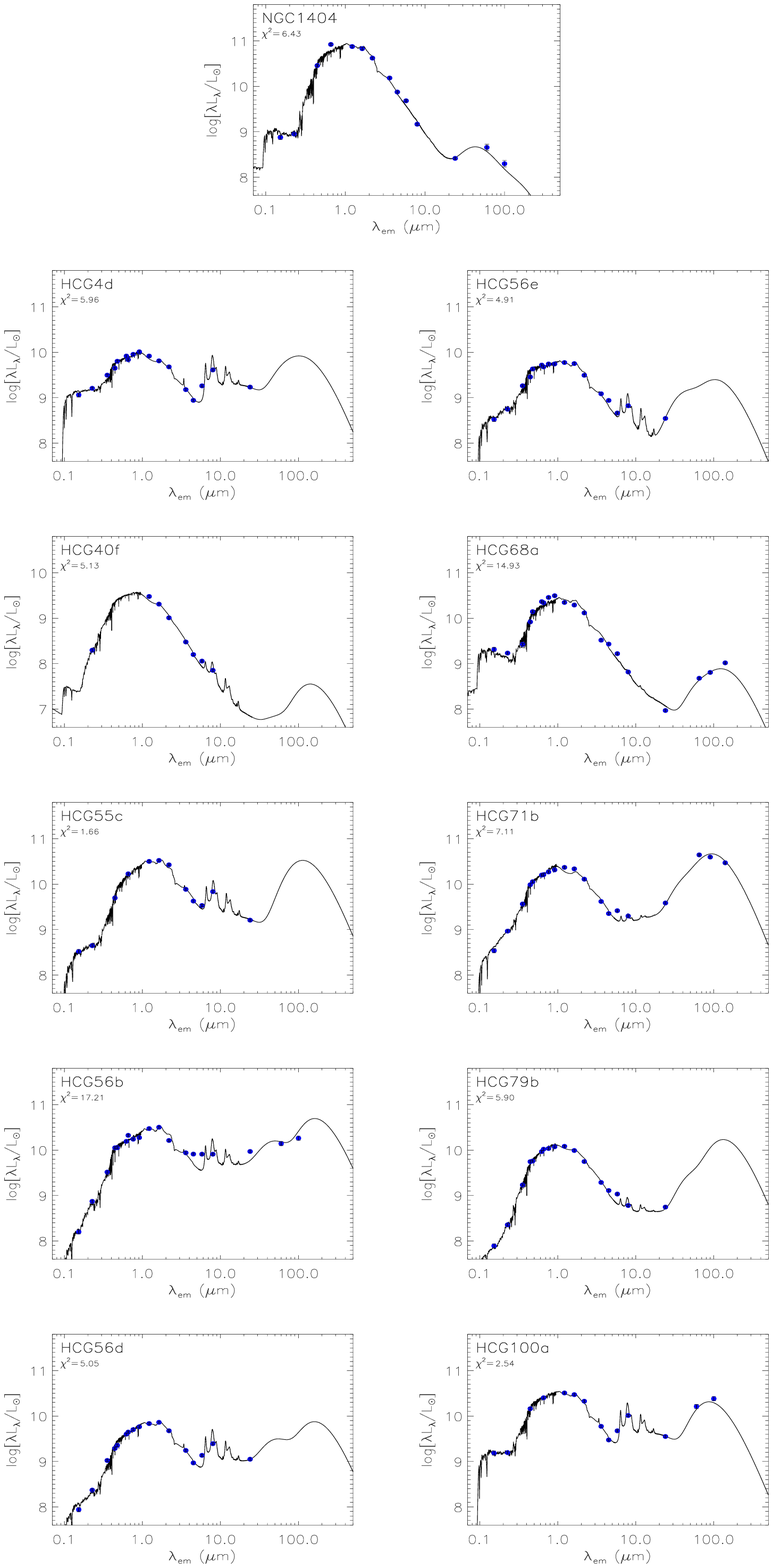}
\caption{We display the observed SEDs (blue points) along with the da Cunha model fits (in black) 
of the 10 peculiar early-type galaxies mentioned in the Appendix. At the top panel we 
present the SED of a typical field elliptical galaxy, NGC1404, for comparison. 
The name of each HCG galaxy along with the $\chi^{2}$ of the model fit to the data is presented.}
\label{fig:appendixsed}
\end{center}
\end{figure}


\end{document}